\documentclass[fleqn,10pt,lineno]{wlpeerj} 
\usepackage{amsmath}
\usepackage{amssymb}
\usepackage{amsfonts}
\usepackage{multirow}
\usepackage{amsthm}

\newtheorem{thm}{Theorem}
\newtheorem{lmma}{Lemma}

\theoremstyle{definition}

\newtheorem{defn}{Definition}

\newcommand{\argmin}{\operatornamewithlimits{argmin}}
\newcommand{\argmax}{\operatornamewithlimits{argmax}}
\newcommand{\mini}{\operatornamewithlimits{min}}
\newcommand{\maxi}{\operatornamewithlimits{max}}
\newcommand \prox {\mathrm{prox}}

\title{Fundamentals of cone regression}

\author[1]{Mariella Dimiccoli}
\affil[1]{University of Barcelona (UB) and Computer Vision Center (CVC)}


\begin{abstract}
Cone regression is a particular case of quadratic programming that minimizes a weighted sum of squared residuals under a set of linear inequality constraints. Several important statistical problems such as isotonic, concave regression or ANOVA under partial orderings, just to name a few, can be considered as particular instances of the cone regression problem. Given its relevance in Statistics, this paper aims to address the fundamentals of cone regression from a theoretical and practical point of view. Several formulations of the cone regression problem are considered and, focusing on the particular case of concave regression as example, several algorithms are analyzed and compared both qualitatively and quantitatively through numerical simulations. Several improvements to enhance numerical stability and bound the computational cost are proposed. For each analyzed algorithm, the pseudo-code and its corresponding code in Scilab are provided. The results from this study demonstrate that the choice of the optimization approach strongly impacts the numerical performances.  It is also shown that methods are not currently available to solve efficiently cone regression problems with large dimension (more than many thousands of points). We suggest further research to fill this gap by exploiting and adapting classical multi-scale strategy to compute an approximate solution.\par
\end{abstract}

\begin{document}

\flushbottom
\maketitle
\thispagestyle{empty}

\section{Introduction}
\label{sec:intro}
Cone regression analysis is a valuable alternative to more traditional parametric-regression models, in all cases where the functional relationships between the response (dependent) and the explanatory (independent) variables is unknown and nonlinear and the constraints are a set of linear inequalities. Several important statistical problems including isotonic, concave and constrained spline regression, or ANOVA under partial orderings can be seen as particular instances of the more general cone regression problem. Cone regression admits several formulation approaches and implementation strategies, whose choice severely impacts numerical performances. However, due to the little exposure to the topics of optimization theory in modern-day Statistics, many optimization and numerical approaches are commonly ignored by statisticians. This paper is a contribution to fill this gap in the literature, addressing the fundamentals of cone regression from a theoretical and practical point of view. With the goal of going in deep with comparisons and numerical issues, we focus in particular on the concave regression problem. In spite of its theoretical simplicity, since the number of constraints increases linearly with the data size,  concave regression offers a good basis to discuss the fundamentals  of cone regression and related numerical issues.

The problem of concave regression is to estimate a regression function subject to concavity constraints represented by a set of linear inequalities.  Brought to the attention of the scientific community by micro-economists interested in estimating production function \cite{Hildreth1954Point,Dent1973Note,Holloway1979Estimation}, the problem of concave regression arises not only in the field of micro-economy (indirect utility, production or cost functions, Laffer curve) but also in medicine (dose response experiments) and biology (growth curves, hazard and failure rate in survival analysis). First addressed by Hildreth in 1954 \cite{Hildreth1954Point}, the search for  efficient methods for solving large concave regression problems is still an open issue nowadays. This may appear quite surprising considering the noticeable advances in convex optimization since then, but it can be probably understood when considering that most efforts have been devoted to theoretical issues such as generalizations and convergence while comparatively little attention has been paid to the issues of efficiency and numerical performances in practice \cite{Perkins2003Convergence,Gould2008How,Censor2009Effectiveness}. 

In this paper, we formulate the cone regression problem by different optimization approaches, we highlight similarities and difference between the various algorithms passed in review, we propose  several improvements to enhance stability and to bound the computational cost and we estimate the expected performance of available algorithms, establishing in particular which is the most competitive technique for solving large instances of the problem. Finally, in the light of this study, we give recommendations for further research.

In  section \ref{sec:statement}, we state formally the problem of cone regression also introducing some basic notations and results that will be used thoroughly. 
In section \ref{sec:StateOfArt} we survey the state of the art distinguishing between the class of algorithms with asymptotic convergence and the class of algorithms with time finite convergence. In section \ref{sec:experiments} we make a numerical comparison of performances and finally, in section \ref{sec:conclusions}, we draw some concluding remarks.

\section{Statement of the problem, basic notations and basic facts}
\label{sec:statement}

The aim of a regression analysis is to produce a reasonable analysis to the unknown response function $f$, which can be modeled as 
\begin{equation}
y = f(z) + \epsilon
\label{regressionModel} 
\end{equation}
where $z \in \mathcal{R}$ is the explanatory (dependent) variable, $y \in \mathcal{R}^d$ is the response (independent) random variable, and $\epsilon$ is an error term, which is usually assumed to be  a mean zero random variable.
Typically, one has observations on $y$ and $z$ for $n$ selected values of $z$.  For each level of input, say $z_i$, there may be several trials and corresponding observations of output $y_i$.
Let $T_i$ be the number of trials at level of input $z_i$ and let $y_{it}$ be the observed output for the $t-$trial at this level. Than we have
\begin{equation}
y_{it} = f(z_i) + \epsilon_{it}, \quad i=1,...,n \quad t=1,...,T_i 
\end{equation}

Inference about the response function may be drawn by assuming that the function $f(z)$ can be approximated by some given algebraic form with several unknown parameters to be estimated from the data. However, the difficulty with this procedure is that the inferences often depend critically upon the algebraic form chosen. Alternatively, one may know some properties of the relation being studied but does not have sufficient information to put the relation into any simple parametric form. In this case, a nonparametric approach is more appropiated. 
Let $x_i$ be the expected value of output at input level $z_i$: 
\begin{equation}
x_i = f(z_i) \quad i=1,...,n 
\end{equation}
Estimates of $x_i$ can be derived by the method of maximum likelihood, or by the method of least squares or other formulations.
If there were no a priori restriction on $f$, than the maximum likelihood estimation of $x_i$ would just be the mean of observed output for the level of input $z_i$, that is 
\begin{equation}
\tilde{x}_i = \frac{1}{T_i} \sum_{T_i}^{i=1} y_{it}\quad i=1,...,n 
\end{equation}
Instead, since in the cone regression problem the known property of the regression function $f$ can be expressed by a set of linear inequalities, to obtain the maximum likelihood estimates, the likelihood function should be maximized subject to the linear inequality constraints.



Formally, given a dataset of $n$ dependent variables represented by the vectors $w, y \in \mathcal{R}^n$, corresponding to the independent variable values $z_1 < z_2 <...< z_n$, 
 the problem of cone regression is to estimate the closest function to the dataset via a least squares regression subject to a set of linear inequality constraints  by solving
\begin{equation} 
\label{regression}
\hat x = \argmin_{\lbrace  x'' \leq 0 \rbrace} \|x-y\|^2_{2,w} 
\end{equation}
$$\mathrm{with}\quad
\|x-y\|^2_{2,w} =   \sum_{i=1}^n w_i(y_i - x_i)^2$$

Denoting by $\mathcal{K}_i$  those vectors that satisfy the linear inequality constraints for a fixed $i$, then $\mathcal{K}_i\neq \varnothing$ is a closed convex  set in $\mathcal{R}^n$ and the feasibility set  $\mathcal{K}$  can be written as the nonempty intersection of a family of closed subsets $\mathcal{K}_i \subset \mathcal{R}^n$. Being the intersection of closed convex sets, the set $\mathcal{K}$ is also a closed convex set. More precisely, since each $\mathcal{K}_i$ is an half-space which contains the origin, the feasibility set $\mathcal{K}$ is a convex polyhedral cone.
In matrix form,  $\mathcal{K}$ can  be  written as $\mathcal{K} = \lbrace x: Ax \leq 0 \rbrace$. In the case of concave regression $A \in \mathcal{R}^{m\times n}$ with $m=n-2$ is a matrix such that each row $A_i$  represents a concave two-piece linear function with a negative second difference at $x_{i+1}$ only and the linear inequalities are as follows
\begin{equation}
\label{eq:constraint}
\frac{x_{i+2}-x_{i+1}}{z_{i+2}-z_{i+1}} - \frac{x_{i+1}-x_{i}}{z_{i+1}-z_{i}} \leq 0,\hspace{0.5cm} i=1,...,n-2
\end{equation}

In the following, we give alternative formulations of the cone regression problem that rest on optimization theory.

\subsection{Convex quadratic programming (CQP) formulation}
\subsubsection{Primal formulation}
The problem (\ref{regression}) is to find the point $\hat{x}$ in the cone $\mathcal{K}$ that is closest to $y$. The solution is found at the orthogonal projection of $y$ onto  $\mathcal{K}$, written as $\Pi(y|\mathcal{K})$ using the metric $\|\cdot\|_{2,w}$, represented by the symmetric positive definite matrix $W$.
\begin{equation} 
\label{eq:primalCQP}
\hat x = \Pi(y|\mathcal{K}) = \argmin_{\lbrace Ax \leq 0 \rbrace} (y-x)^T W (y-x) 
\end{equation}
For the problem (\ref{eq:primalCQP}), the matrix $W$ is diagonal with element $w_i$ on the diagonal. In practice, if $y_i$ is the mean value measured at $z_i$, than $w_i$ corresponds to the size of the sample at $z_i$.
Since $K$ is a closed, convex and non empy set on the Hilbert space $\mathcal{R}^n$, it is a \textit{set of Chebyshev}, that is the projection exists and it is unique.  
\begin{figure}
   \begin{center}
	\includegraphics[width=6.0cm]{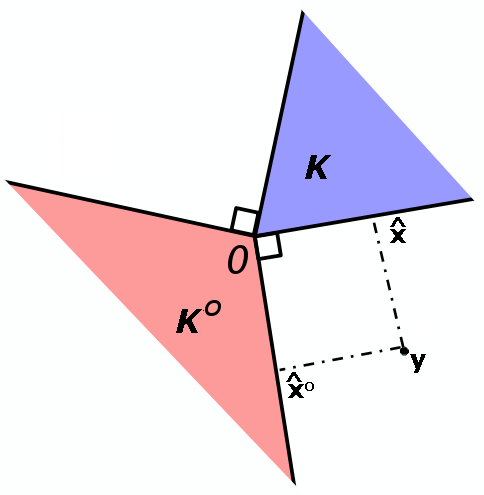}
    \end{center}
\caption{The polar cone $\mathcal{K}^o$ of a given convex cone $\mathcal{K} \subset \mathcal{R}^2$ is given by the set of all vector whose scalar product with vectors of $\mathcal{K}$ is negative. The data point $y$ can be written as the sum of $\hat{x}$, the projection onto the cone $\mathcal{K}$ and $\hat{x}^o$, the projection onto the polar cone $\mathcal{K}^o$.}
       \label{fig:polarcone}
\end{figure}

\subsubsection{Dual formulation}
The dual formulation of problem (\ref{eq:primalCQP}) rests on the Moreau decomposition theorem \cite{Moreau1962Decomposition}, which is a generalization in convex analysis of the orthogonal projection theorem for vectorial sub-spaces.
Central to the Moreau decomposition theorem is the definition of \textit{polar cone} of a given convex cone $\mathcal{K}$, which is given below.
\begin{defn} 
\label{def:polarcone}
The \textit{polar cone} $\mathcal{K}^o$ to any convex cone $\mathcal{K}$ is given by 
\begin{equation} 
\mathcal{K}^o= \lbrace x \in \mathcal{R}^n : \forall k \in \mathcal{K},\langle  k',x \rangle \leq 0 \rbrace
\end{equation} 
\end{defn} 

The Moreau decomposition theorem is as follows.
\begin{thm}
\label{lm:moreau}
Let  $\mathcal{K} \subseteq \mathcal{R}^n$ be a closed convex cone, $\mathcal{K}^o$ its polar cone and $y \in \mathcal{R}^n$ a given point. Then the following assertions are equivalent:
\begin{eqnarray*}
(i) \hspace{0.2cm} \hat{x} =  \argmin_{x \in \mathcal{K}} ||x-y||^2, \hspace{0.1cm} \hat{x}^o = \argmin_{x \in \mathcal{K}^o} ||x-y||^2 \\
(ii) \hspace{0.2cm} \hat{x} \in \mathcal{K}, \hspace{0.2cm}\hat{x}^o \in \mathcal{K}^o, \hspace{0.2cm} \langle \hat{x},\hat{x}^o \rangle =0, \hspace{0.2cm} y = \hat{x} + \hat{x}^o
\end{eqnarray*}
\end{thm}
By relying on this theorem we can alternatively solve problem (\ref{eq:primalCQP}) by first finding the projection on the polar cone $\hat{x}^o$ and then computing the solution to the primal problem as the difference $\hat{x} = y - \hat{x}^o$ (see Fig. \ref{fig:polarcone}). This alternative is attracting since, as it will be clarified below, it implies an analytically simpler form for the constraints. 

Before stating an important Lemma about the relationship between the polar cone and the constraint matrix $A$, let us introduce the definition of \textit{edges} of a  polyhedral convex cone.
\begin{defn} 
\label{def:edgecone}
Let $\mathcal{K}$ be a polyhedral convex cone in $\mathcal{R}^n$, then the vectors $e_i \in \mathcal{R}^n\setminus\{0\}$ are the \textit{edges or generators} of $\mathcal{K}$ if and only if $\mathcal{K}=pos(\{e_i\}) = \{ \sum k_ie_i | k \geq 0\}$. 
\end{defn} 
Intuitively speaking, the edges of a polyhedral convex cone are one-dimensional rays, which always passe through a fixed point (the vertex). 

\begin{lmma}
\label{lm:edgesPolar}
The rows of the matrix $A$ are the edges of the polar cone, that is $\mathcal{K}^o = \lbrace x : x= \sum_{i=1}^m A_i^T a_i,a_i\geq0 \rbrace$.
\end{lmma}
To see that, observe that $\mathcal{K}^o=\lbrace \sum_{i=1}^m a_i A^T_i, a_i \geq 0 \rbrace$ is polar to  $\mathcal{K}$ since

$\forall x \in \mathcal{K}$,$\forall \rho \in \mathcal{K}^o$: $\langle \rho, x \rangle = \lbrace \sum_{i=1}^m a_i \langle A^T_i,x \rangle \leq 0\rbrace$, which is the definition of polar cone of $\mathcal{K}$. Conversely, $\mathcal{K}$ is polar to $\mathcal{K}^o$ since: $\mathcal{K} = (\mathcal{K}^o)^o = \mathcal{K}$.

By relying on this results Khun-Tucker \cite{kuhn1951nonlinear} proved the following theorem:
\begin{thm}
The primal constrained quadratic minimization problem (\ref{eq:primalCQP}) is equivalent to the dual problem
\begin{equation} 
\label{eq:dualCQP}
\hat \lambda =  \argmin_{ \lambda \geq 0 } (y-A^T\lambda)^T W (y-A^T\lambda) 
\end{equation}
Denoting by $\hat{\lambda}$ the solution to the dual problem, the solution to the primal problem is $\hat{x} = y - A^T\hat{\lambda}$.
\end{thm}

As it can be observed, in the dual formulation each element of the vector $\lambda$ must satisfy a single positivity constraint.

Goldman \cite{Goldman1993Nonparametric} noticed that dual problem can be also view as a minimum distance problem in the same parameter space as the primal problem.
\begin{equation} 
\label{eq:reparametrizeddualCQP}
\hat{x} = \argmin_{x \in \mathcal{C}}||x||^2 
\end{equation}
where $\mathcal{C} = \lbrace x| x = y - A^T\lambda, \lambda \geq 0 \rbrace$ is a rotation of the dual cone with its vertex translated to $y$. $\hat{x}$ also solves the re-parametrized dual problem.

\subsection{Linear complementarity problem (LCP) formulation}
\label{subsec:LCP}
The CQP (\ref{eq:primalCQP}) can be recasted as a linear complementarity problem (LCP). To see that, let us consider the Lagrangian associated to problem (\ref{eq:primalCQP}).
\begin{equation}
\label{eq:lagrangian}
   L(x,\lambda) =  ||x-y||^2_{2,w} + <\lambda,Ax>  
\end{equation}
where $\lambda \geq 0$ is the vector of dual variables associated to each of the convexity constraints.
By applying the Karush–Kuhn–Tucker (KKT) optimality conditions \cite{kuhn1951nonlinear} to (\ref{eq:lagrangian}), that is 
\begin{eqnarray}
\label{eq:KTT}
\nabla L(\hat{x},\hat{\lambda}) = 0\\
\lambda \geq 0\\
\lambda^TA\hat{x} = 0
\end{eqnarray}
we obtain the equivalent LCP
\begin{eqnarray}
\label{eq:LCP}
w + M\lambda = q\\
w\geq 0, \quad \lambda  \geq 0, \quad w^T\lambda  =0.
\end{eqnarray}
where $w = -A^Tx$, $M= -AA^T$ and $q = -A^Ty$.
Note that by dropping the constant term from the Lagrangian and dividing it by $2$: $L(x,\lambda) =  \frac{1}{2}xx^T-y^Tx + \lambda^TAx$. Therefore: $\nabla L(x, \lambda) = x^T -y^T + \lambda^TA = 0$. By taking the transpose and multiplying for $-A$: $-Ax + (-AA^T)\lambda = -A^Ty$.
This LCP has a unique complementary solution. Denoting by $(\hat{w},\hat{\lambda})$ its solution, $\hat{\lambda}$ is the optimal solution of the dual problem (\ref{eq:dualCQP}).

The condition $w^T\lambda =0$ is called complementarity condition and the way in which it is dealt with determines if the optimization algorithm belongs to the class of \textit{interior point methods} that will be introduced in section \ref{subsubsec:PDinteriorPointMethods} or to the class of  \textit{active set  methods} that will be detailed in section \ref{subsec:finiteconvergence}.

\subsection{Proximal formulation}
The CQP (\ref{eq:primalCQP}) can be solved by using a proximity operator \cite{Moreau1962Fonctions,Moreau1963Proprietees}. Proximity operators are used to solve problems of the form
\begin{equation}
 \argmin_{x \in \mathcal{R}^n}  f_1(x) + f_2(x)...+ f_m(x)
\end{equation}
where $f_1,f_2,...,f_m$ are convex functions from $\mathcal{R}^n$ to $]-\infty,+\infty]$, that are not necessarily differentiable.
Each $f_i$ is treated through its proximity operator which is defined as follows.
\begin{defn}
Let $\Gamma_0(\mathcal{R}^n)$ be the class of lower semicontinuous convex functions from $\mathcal{R}^n$ to $]-\infty,+\infty]$ such that their domain, denoted by $dom(f)$ is not the empty set. Let $f$ be a function $f \in \Gamma_0(\mathcal{R}^n)$, then the \textit{proximity operator} of $f$ is $prox_f(x): \mathcal{R}^n \rightarrow \mathcal{R}^n$ such that
\begin{equation}
 \forall x \in \mathcal{R}^n, \quad prox_f(y) = \argmin_{x\in \mathcal{R}^n}  f(x) + \frac{1}{2}||x-y||^2 
\end{equation}
\end{defn}
The proximity operator is characterized by the property 
\begin{equation}
\forall (x,p) \in \mathcal{R}^n \times \mathcal{R}^n, \quad p = prox_f(y)  \quad\Longleftrightarrow\quad  y-p \in \partial f(p),
\end{equation}
where $\partial f: \mathcal{R}^n \rightarrow 2^{\mathcal{R}^n}$ is the subdifferential of $f$.
\begin{equation}
\partial f = \Big \lbrace u \in \mathcal{R}^n : \forall y \in \mathcal{R}^n: (y-p)^Tu + f(p) \leq f(y)) \Big \rbrace
\end{equation}
The proximity operator of a convex function is a generalization of the projection operator onto a closed convex set $\mathcal{C}$. To see that let us consider the indicator function of $\mathcal{C}$
\begin{center}
\[\imath_\mathcal{C}(x)= \left\{
\begin{aligned}
	&0\text{ if }x\in\mathcal{C}\\
	&+\infty\text{ if }x\not\in\mathcal{C}
\end{aligned}
\right.\]
\end{center}
By using the fact that minimizing $J(x)$ over $\mathcal{C}$ is equivalent to minimizing $J(x) + \imath_\mathcal{C} (x)$ over $\mathcal{R}^n$
\begin{center}
 $\argmin\limits_{\substack{x\in \mathcal{C}}} J(x) ~~=~~ \argmin\limits_{\substack{x\in \mathcal{R}^n}} \{J(x) + \imath_\mathcal{C} (x) \}$
\end{center}
it results that  $\prox_{\imath_\mathcal{C}} = \Pi(\cdot|\mathcal{C})$.

The solution to problem (\ref{eq:primalCQP}) can be therefore understood as the proximity operator over $\mathcal{K}$
\begin{align}
\label{eq:proximalDual1}
\hat{x}= \argmin\limits_{\substack{x\in \mathcal{K}}}||x-y||^2 
= &\argmin\limits_{\substack{x\in \mathcal{R}^n}}\{||x-y||^2 + \imath_{\mathcal{K}}(x)\}
\\
= & \hspace{0.1cm}\prox_{\imath_{\mathcal{K}}}y 
\end{align}

Alternatively, using the fact that $(\mathcal{K}_1 \cap ... \cap \mathcal{K}_m)^o = \mathcal{K}^o_1 + ... + \mathcal{K}^o_m$, the dual problem (\ref{eq:primalCQP}) can be seen as the proximity operator over the sum of $m$ indicator functions of the convex sets $\mathcal{K}^o_i$:
\begin{align}
\label{eq:proximalDual}
\hat{x}^o= \argmin\limits_{\substack{x\in \sum_{i=1}^m\mathcal{K}^o_i}}||x-y||^2 
= &\argmin\limits_{\substack{x\in \mathcal{R}^n}}\{||x-y||^2 + \sum_{i=1}^m\imath_{\mathcal{K}^o_i}(x)\}
\\
= & \hspace{0.1cm}\prox_{\sum_{i=1}^m\imath_{\mathcal{K}^o_i}}y 
\end{align}

Intuitively, the base operation of a proximal algorithm is evaluating the proximal operator of a function, which itself involves solving a small convex optimization problem. These subproblems often admit closed form solutions or can be solved very quickly with standard or simple specialized methods. 

\section{State of the art}
\label{sec:StateOfArt}
In this section, we review existing algorithms for solving the cone regression problem. The algorithmic approaches, and in turn their numerical performances, strongly depend on the choice of the problem formulation. All existing methods are iterative and they can attain the optimal solution since $\mathcal{K}$ is a Chebyshev set and therefore the optimal solution must exist in this closed set. However, in terms of their numerical perfomances they can be classified into two broad classes: the class of methods never or in very simple cases attain the optimal solution \cite{Hildreth1954Point,Dykstra1983Algorithm} and those of methods that converge to the optimal solution in a finite number of steps \cite{Wilhelmsen1976Nearest,Pshenichny1978Numerical,Wu1982Some,Fraser1989Mixed,Meyer1999Extension,Meyer2013Simple,Murty1982Critical,Liu2011Active}. As it will be  clarified in the following,  methods with asymptotic convergence rest on the properties of the sub-gradient or more  in general of proximity operators and act by finding the solution as the limit of a sequence of successive approximations. They are  typically derived from the primal, the dual or from the proximal formulation.  Methods with finite-time convergence exploit the geometric properties of polyhedral convex cones and find the exact solution as non-negative linear combination of functions, forming a basis in a specified finite dimensional space. They are typically derived from the linear complementarity problem formulation.


\subsection{Algorithms with asymptotic convergence}

This section includes algorithms based on the primal formulation such as Least Squares in a Product Space (section \ref{subsubsec:LSPS}, algorithms based on the dual formulation such as the Uzawa's method (section \ref{subsubsec:uzawa}) and Hildret's method  (section \ref{subsubsec:hildret}), algorithms that solve the dual problem simultaneously with the primal problem such as the Dykstra's alternating projection method (section \ref{subsubsec:dykstra}) and algorithms based on the proximal formulation such as Alternating Direction Method of Multipliers  (section \ref{subsubsec:ADMM}).

\subsubsection{Least squares in a product space (LSPS)}
\label{subsubsec:LSPS}
Since the Euclidean space $\mathcal{R}^n$ equipped with the dot product is an Hilbert space $\mathcal{H}$, the problem (\ref{eq:primalCQP}) can be recasted in the $m-$fold Cartesian product of $\mathcal{H}$, say  $\mathcal{H}^m$ \cite{Pierra1984Decomposition}. Let  $\mathcal{K}^m$ be the Cartesian product of the sets $(K_i)_{i \in I}$, i.e., the closed convex set 
$\mathcal{K}^m = \times_{i \in I} \mathcal{K}_i= \lbrace x \in \mathcal{H} : \forall i \in I: x_i \in \mathcal{K}_i \rbrace$ and let $D$ be the diagonal vector subspace, i.e. $D = \lbrace (x,...,x) \in \mathcal{H}^m : x \in \mathcal{H} \rbrace$. 
Then, the CQP  (\ref{eq:primalCQP}) is equivalent to 
\begin{equation}
\label{eq:lsps}
\argmin_{\lbrace x  \in \mathcal{K}^m \cap D \rbrace} \|x-\bar{y}\|^2_{2,w} 
\end{equation}
where $\bar{y} = (y,...,y)$.
Using this strategy, the problem of projecting onto the intersection of $m$ convex sets is reduced to the problem of projecting, in an higher dimensional space, onto only two convex sets, one of which is a simple vector subspace. Geometrically, this is equivalent to find a point in $D$  which is at minimum distance from $\mathcal{K}^m$.  This point can be obtained iteratively by 
\begin{equation}
 x_{k+1} = x_k + \lambda_k(P_D \circ P_{\mathcal{K}}^m(x_k) - x_k)
\end{equation}
The advantage of this strategy is in that it allows to speed up the convergence since a bigger (than 2, which is the upper bound for Féjer sequences \cite{Eremin1969Fejer}) relaxation interval can be allowed.

\subsubsection{Uzawa method}
\label{subsubsec:uzawa}
A classical method to solve a convex minimization problem subject to inequality constraints is the Uzawa method \cite{Arrow1958Studies}, which search directly for the saddle point of the Lagrangian  (\ref{eq:lagrangian}).
In fact, if the Lagrangian $L(x,\lambda)$ admits a saddle point, say $(\hat{x}, \hat{\lambda})$, then the duality gap $\delta = \mini_{x \in \mathcal{K}}\maxi_{\lambda \in \mathcal{R}^+}L(x,\lambda) - \maxi_{\lambda \in \mathcal{R}^+}\mini_{x \in \mathcal{K}}L(x,\lambda)$ is null and $\hat{x}$ is a critical point of the Lagrangian.
Since the dual function $H(\lambda) = \argmin_{x \in \mathcal{K}} L(x,\lambda)$ is differentiable, it can be minimized explicitly  by using the gradient descent method. Therefore the Uzawa method alternates a minimization step over $\mathcal{R}^n$ with respect to $x$ with $\lambda$ fixed and a maximization step with respect to  $\lambda$ onto $\mathcal{R}^+$, with $x$ fixed.
The algorithmic parameter $\rho > 0 $ can be fixed to optimize convergence by relying on theoretical considerations.
Therefore the CQP (\ref{eq:dualCQP}) is equivalent to find
\begin{equation}
\label{eq:uzawa}
\hat{x} =\argmin_x \argmax_{\mu \geq 0} L(x, \mu) =  ||x-y||^2 + <\mu,Ax> 
\end{equation}

\subsubsection{Hildreth's algorithm}
\label{subsubsec:hildret}
Hildreth \cite{Hildreth1954Point}  proposed to apply the Gauss Seidel algorithm \cite{Kahan1958GaussSeidel} to the dual problem (\ref{eq:dualCQP}). 
A single cycle of the Hildreth's algorithm consists in updating each element of $\lambda$ sequentially in an arbitrary fixed order. Therefore each cycle consists of $m$ steps, each of which corresponds to a projection onto the cone $ \mathcal{K}_i$,$i=1,...,m$. The algorithm gives rise to a sequence of points, each of one differs from the preceding in exactly one coordinate. At the cycle $k+1$, the $ \lambda_{i}^{k+1}$ is used in the estimation of the point $\lambda_{i+1}^{k+1}$ so that the best available estimations are used for computing each variable.
The convergence of the Gauss Seidel algorithm is guaranteed only if the matrix $A$ is full row rank, so that there are not redundancies among the inequality restrictions, and it is guaranteed independently of the initial point $\lambda^0$ only if $A$ is positive definite and symmetric. 
The algorithm is sensitive to the normalization as well as to the order of the projections.

\subsubsection{Primal-dual interior point methods}
\label{subsubsec:PDinteriorPointMethods}
First introduced by Karmakar in 1984 \cite{Karmarkar1984New}, primal-dual interior point methods act by perturbing the complementarity condition $w^T\lambda= 0$ in the LCP formulation \ref{eq:LCP} and replacing  with $w^T\lambda= \mu$. The partition of vectors $w$ and $\lambda$ into zero and nonzero elements is gradually revealed as the algorithm progresses by forcing a reduction of $\mu$. All iterates satisfy the inequality constraints strictly. The solution is approached from either the interior or exterior of the feasible region but never lie on the boundary of this region. 
Let the function $F(x, \lambda, w)$ be such that the roots of this function are solutions to the first and the last optimality conditions in \ref{eq:LCP}. 

\vspace{0.5cm}
$F_{\mu}(x, \lambda, w) = $$\begin{pmatrix}
        w - A^Tx\\[0.3em]
  x^T - y^T + \lambda^tA\\[0.3em]
         w^T\lambda  - \mu e\\[0.3em]
     \end{pmatrix}$
\vspace{0.3cm}

The perturbed complementarity condition introduces a nonlinearity, therefore for each fixed $\mu > 0$ a system of nonlinear equations should be solved.
The nonlinear system is typically solved by using a Newton-like algorithm \cite{ben1966newton}. Each iteration of the Newton's method finds a search direction from the current iterate $(x_k, \lambda_k, s_k)$ and it is computationally expensive but can make significant progress towards the solution.  For instance in barrier methods, which are the most efficent of the family, this is achieved by using a penalizing term, called barrier function, for violations of constraints whose value on a point increases to infinity as the point approaches the boundary of the feasible region. Interior point methods must be initialized at an interior point, or else the barrier function is undefined. 
The interested reader is referred to  \cite{Singh2002InteriorPoint} for further information about interior point methods.

\subsubsection{Dykstra's algorithm}
\label{subsubsec:dykstra}
In $1983$, Dykstra \cite{Dykstra1983Algorithm} proposed a generalization of the Hildreth's procedure applicable to the case of constraints corresponding to more general convex cones than polyhedral convex ones. The Dykstra's algorithm is based on the idea, before suggested by Von Neumann \cite{vonNeumann1950Functional} to the case of subspaces, of computing the projection onto the intersection of convex sets by relying on the solution of the simpler problem of projecting onto the individual sets. In the case of concave regression the projection onto a single convex set $\mathcal{K}_i$ involves only three points and, if the constraint is not satisfied, it corresponds to the straight line fitting the points $y_i$,$y_{i+1}$,$y_{i+2}$.

Dykstra's algorithm iterates by passing sequentially over the individual sets and projects onto each one a deflected version of the previous iterate. More precisely, before projecting onto the cone $\mathcal{K}_i$ during the $(k+1)-$th cycle, the residuum obtained when projecting onto $\mathcal{K}_i$ at the previous $k-$th cycle, say $R_i^k$ is removed and a new residuum associated to the cone $\mathcal{K}_i$, say $R_i^{k+1}$ is computed after the projection. 
In practice, each $x^k_i$ is the projection of $y + R^k_1 +...+R^k_{i-1}+R^k_{i+1}+...+R^k_m$ onto $\mathcal{K}_i$, where $R^k_i= x^k_i-(y + R_1^k + R^k_{i-1} +R^{k-1}_{i+1}+...+R^{k-1}_m)$.

If each $\mathcal{K}_i$ is a subspace, then at each new cycle $k+1$ the residuum $-R_i^k$ of the projection onto each convex cone $\mathcal{K}_i$ is of course the projection of $x^{k}$ onto the cone $\mathcal{K}_i^o$. Therefore, the Dykstra's procedure for subspaces reduces to exactly the cyclic, iterated projections of von Neumann. In this case, for $k \rightarrow \infty$ , the sum of the residua over the cones $\mathcal{K}_i^k$ approximates the projection of $y$ onto the polar cone $\mathcal{K}^o$ and therefore, for the Moreau decomposition theorem, $x^{k}$ approximates the projection onto $\mathcal{K}$.

However, if the $\mathcal{K}_i$ are not subspaces, $\Pi(\cdot|\mathcal{K}_i)$ is not a linear operator and then the von Neumann algorithm does not necessarily converge.

The Dykstra's algorithm can also be interpreted as a variation of the Douglas–Rachford splitting method applied to the dual proximal formulation (\ref{eq:proximalDual}).


The seminal works of Hildreth and Dykstra have inspired many studies mostly devoted to theoretical investigations about their behavior in a Hilbert space \cite{BoyleMethod,Varian1984Nonparametric}, about its convergence \cite{Iusem1991Convergence,Crombez1995Finding}, about its relation to other methods \cite{Gaffke1989Cyclic,Bauschke1994Method} and about its interpretation in more general frameworks, such as the proximal splitting methods \cite{Combettes2011Proximal}. Han \cite{Han1988Successive}, as well as Iusem and De Pierro \cite{Iusem1991Convergence}, showed that in the polyhedral case, the method of Dysktra becomes the Hildreth's algorithm and therefore it has the same geometric interpretation of Gauss Seidel to the dual problem. Gaffke and Mathar (1989) \cite{Gaffke1989Cyclic} showed the relation of the Dysktra algorithm to the method of component-wise cyclic minimization over a product space, also proposing a fully simultaneous Dykstra algorithm. The only works devoted to give some insight about a more efficient implementation are the ones of Ruud. Goldman and Ruud \cite{Goldman1993Nonparametric}($1993$) generalized the method of Hildreth showing that there is not need to restrict the iterations to one element of $\lambda$ at a time: one can optimize over subsets or/and change the order in which the elements are taken. This observation is important for the speed of convergence since the slow speed can be understood as a symptom of near multicollinearity among restrictions. Because the intermediate projections are so close to one another, the algorithm makes small incremental steps towards the solution. They also remarked that Dykstra uses a parametrization in the primal parameter space, that causes numerical round off errors in the variable residuum. These round off errors cumulate so that the fitted value does not satisfy the constraints of the dual problem. It would be better to use a parametrization on the dual so that the contraints in the dual would be satisfied at each iteration. 
Later, Ruud \cite{Ruud1997Restricted} proved that the contraction property of the proposed generalizations rests solely on the requirement that every constraints appears in at least one subproblem of an iteration. As one approach the solution, constraints that are satisfied at the solution are eliminated. To remove satisfied constraints would accelerate the Hildreth procedure. The authors propose to reduce the set of active constraints, that is the constraints satisfied as an equation at the corresponding points, by removing as many constraints as possible through periodic optimization over all positive elements of $\lambda$.

\subsubsection{Alternating Direction Method of Multipliers (ADMM)}
\label{subsubsec:ADMM}
ADMM is an augmented Lagrangian technique \cite{Hestenes1969Multiplier,Powe69a} which can be applied to problems of the form 
\begin{equation}
\label{eq:admm}
Find \argmin_{z\in \mathcal{R}^m, Ax=z, z\leq 0} ||y-x||^2 + g(Ax)
\end{equation}
where the matrix $A$ is assumed to be irreducible ($AA^T= vI, v>0$) and the intersection of the relative interiors of the domains of the two functions is assumed to be not empty ($ ri \hspace{0.2cm}dom(g) \cap ri\hspace{0.2cm} dom(f) \neq \varnothing$).
ADMM  minimizes the augmented Lagrangian $\mathcal{L}$ over the two variables of the problems, say $x$ and $z$, first $x$ with $z$ fixed, then over $z$ with $x$ fixed, and then applying a proximal maximization step with respect to the Lagrange multiplier $\lambda$.
The augmented Lagrangian of index $\gamma \in [0, \infty]$  is
\begin{equation}
 \mathcal{L}(x,z,y) = f(x) + g(z) + \frac{1}{\gamma} \lambda^T(Ax - z) + \frac{1}{2\gamma} ||Ax - z||^2
\end{equation}
where $f(x) = ||y-x||^2$.
Denoting by $prox_f^A$ the proximal operator which maps a point $z \in \mathcal{R}^n$ to the unique minimizer of $f(x) + ||Ax-z||^2$ and denoting by $prox_g = prox_{f \circ A}$ the implementation detailed in the Appendix is obtained.

The ADMM method rests on the proximal formulation \ref{eq:proximalDual1}. Indeed, it can be viewed as an application of the Douglas-Rachford splitting algorithm \cite{Eckstein1992DouglasRachford}.




\subsection{Algorithms with time-finite convergence}
\label{subsec:finiteconvergence}

All algorithms that will be reviewed in this section are \textit{active set methods} resting on the LCP formulation \ref{eq:LCP}. Active set methods work by choosing a subset of indices $j \in \tilde{J} \subset J = \{1, . . . , n\}$ such that $w_j$ is allowed to be non-zero and forcing the corresponding $\lambda_j$ to be zero, while the remaining indices $j \in J \setminus \tilde{J}$ force $w_j$ to be zero and allow $\lambda_j$ to take nonzero values.
In this section we will review active set algorithm suach as the mixed primal-dual basis algorithm (section \ref{subsubsec:fraser}), the critical index algorithm (section \ref{subsec:critical}) and the Meyer's algorithm (section \ref{subsubsec:hinge}). 


Before detailing the algorithms, we introduce some definitions and basic results about the geometry of polyhedral convex cones, on which are based the algorithms presented in  this section. For further details the reader is referred to \cite{silvapulle2011constrained}.

\subsubsection{Properties of polyhedral convex cones with $m\leq n$}
\label{subsec:properties}

Lemma \ref{lm:edgesPolar} establishes the relationship between the constraint matrix $A$ and the edges of the polar cone $\lbrace \gamma^i, i= 1,...,m\rbrace$, namely $A^T= [\gamma^1, ...,\gamma^m]$.
We would like now to determine the edges of the constraint cone $\mathcal{K}$.

Let the vectors $\lbrace \gamma^{m+1},.., \gamma^{n}\rbrace$  vectors orthogonal to $\lbrace\gamma^i, i = 1,..,m\rbrace$ and orthonormal to each other so that the set $\lbrace \gamma^i, i = 1,..,n\rbrace$ forms a basis for $\mathcal{R}^n$. By defining the dual basis of the basis $\lbrace \gamma^i, i = 1,..,n\rbrace$ as the set of vectors $\lbrace \beta^i, i= 1,...,n \rbrace$ that verify the
relationship
\begin{equation}
\label{eq:dualbasis}
 (\beta^i)^T \gamma^j = \left\{\begin{array}{ccll} -1 \hspace{0.5cm} i = j \\ 0 \hspace{0.5cm} i \neq j \end{array} \right. 
\end{equation}
the constraint cone $\mathcal{K} = \lbrace x: Ax \leq 0\rbrace$ can be equivalently written as 
\begin{equation}
\label{constraintCone2}
\mathcal{K} = \Big  \lbrace x: x = \sum_{i=1}^m b_i\beta^i + \sum_{i=m+1}^n b_i\beta^i, b_i \geq 0, i = 1,...,m \Big  \rbrace 
\end{equation}

To see that let $B = [\beta^1,...,\beta^n]$ and $C= [\gamma^1,...,\gamma^n]$. Then $Ax$ are the first $m$ coordinates of $Cx$. Since $B^TC = -I_n$ by construction, by multiplying both members at left for $B^{-1}$ and at right for $x$, we obtain: $Cx = -B^{-1}x$. Therefore $Cx$ gives the negative coordinates of $x$ in the basis $\lbrace \beta^i, i = 1,...,n \rbrace$. Furthermore, points in $\mathcal{K}$ have their first $m$ coordinates non-negative and can be written as $x = \sum_{i=1}^n b_i\beta^i$, where $b_i \geq 0$ for $i=1,..,m$.

Taking into account Def. \ref{def:edgecone},   Eq. \ref{constraintCone2} established that the vectors $\beta^i$ are the edges of the constrain cone $\mathcal{K}$.

\begin{defn} 
\label{def:facecone}
Let $\mathcal{K}$ be a polyhedral convex cone in $\mathcal{R}^n$, then $F \subseteq \mathcal{K}$ is a \textit{face} of $\mathcal{K}$ if and only if $F$ is the intersection of $\mathcal{K}$ with a supporting hyperplane.
\end{defn} 

A polyhedral convex cone arises as the intersection of a finite number of half-spaces whose defining hyperplanes pass through the origin. The $i-$th row of $A$ is normal to the hyperplane generating the $i-$th closed half-space.

The following Lemma, proved by Rockafellar in 1970 \cite{Rockafellar1970Convex}, establishes a relationship between the first $m$ vectors of the dual basis $\lbrace \beta^i, i= 1,...,m \rbrace$  and the faces of the cone $\mathcal{K}' = \mathcal{K} \cap span(\mathcal{K})$, where $span(\mathcal{K})$  denotes the subspace spanned by the $m$ edges of $\mathcal{K}$.

\begin{lmma} 
\label{def:Partitionface}
Let  $\mathcal{K} = \lbrace x: Ax \leq 0 \rbrace$, where $A^T = [\gamma_1,...,\gamma_m]$, be  the constraint cone and let $\lbrace \beta^i, i= 1,...,m \rbrace$ be the dual basis of  $\lbrace \gamma^i, i= 1,...,m \rbrace$. Denoting by $span(\mathcal{K})$  the subspace spanned by the $m$ edges of $\mathcal{K}$, let  $\mathcal{K}' = \mathcal{K} \cap span(\mathcal{K}) = \lbrace  x \in \mathcal{R}^n: x =  \sum_{j=1}^m b_j\beta^j, b_j \geq 0 \rbrace$. Then, for $J \subseteq \lbrace {1,...,m} \rbrace$ the faces of $\mathcal{K}'$ are the sets: 
\begin{equation*}
\Big \lbrace  x \in \mathcal{R}^n: x =  \sum_{j \in J}b_j\beta^j, b_j \geq0\Big \rbrace  
\end{equation*}
The set of all relatively open faces 
\begin{equation*}
\mathcal{F}_J = \Big \lbrace  x \in \mathcal{R}^n: x =  \sum_{j \in J}b_j\beta^j, b_j > 0\Big \rbrace  
\end{equation*}
forms a partition of $\mathcal{K}'$.
\end{lmma}

\begin{figure}
   \begin{center}
	\includegraphics[width=6.0cm]{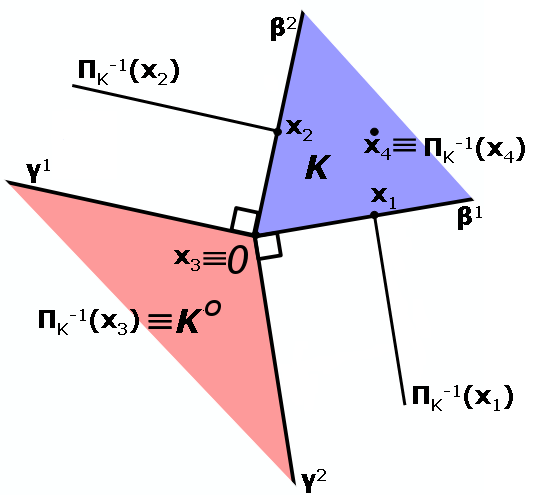}
    \end{center}
\caption{The point $x_1 \in \mathcal{K}$ belongs to the open face $F_J=  \lbrace x \in \mathcal{K}: x = b\beta^1, b >0 \rbrace$, with $J = \lbrace 1\rbrace$. The support cone of  $\mathcal{K}$ at $x_1$ is $\mathcal{L}_{\mathcal{K}}(x_1) = \lbrace x :  {\gamma^2}^Tx \leq 0\rbrace$ and its dual is $\mathcal{L}^o_{\mathcal{K}}(x_1) = \lbrace x : x =  c\gamma^2, c \geq 0\rbrace$. The set of points that project onto $x_1$ is given by the set $\Pi^{-1}_{\mathcal{K}}(x_1) = \lbrace x_1 +  \mathcal{L}^o_{\mathcal{K}}(x_1) \rbrace = \lbrace x_1 + c\gamma^2, c \geq 0 \rbrace$. 
The point $x_3 \in \mathcal{K}$ belongs to the open face $F_J=  \lbrace 0 \rbrace$, with $J = \varnothing$. The support cone of  $\mathcal{K}$ at $x_3$ is $\mathcal{K}$ and its dual is $\mathcal{K}^o$, so that the set of points that project onto $x_3$ is $\lbrace \mathcal{K}^o \rbrace$. 
The point $x_4 \in \mathcal{K}$ belongs to the open face $F_J=  \lbrace x \in \mathcal{K}: x = \sum_{i=1,2}b_i\beta^i, b_i >0 \rbrace$, with $J = \lbrace 1,2\rbrace$.
The support cone of  $\mathcal{K}$ at $x_4$ is the origin and its dual is the origin, so that the set of points that project onto $x_4$ is $\lbrace x_4  \rbrace $. }
       \label{fig:Partition}
\end{figure}

Denoting by $u$ and $v$  the  projections of $y$ onto $span(\mathcal{K})$ and $span(\lbrace\gamma^{m+1},...,\gamma^{n}\rbrace)$ respectively, being the two subspaces orthogonal,  $y$ can be written as $y = u + v$.  Since $v$  can be easily computed as $v =  X(X^TX)^{-1}X^T y$ with $X = [\gamma^{m+1},...,\gamma^{n}]$, the problem (\ref{eq:primalCQP}) reduces to find the projection of $u$ onto $\mathcal{K}'$.

The next Lemma, proved by Zarantonello \cite{Zara71a} focuses on the set of points in $span(\mathcal{K})$ projecting on a given point $x \in \mathcal{K}'$, say $\Pi^{-1}_{\mathcal{K}'}(x)$. Before stating it we need to define the \textit{support cone} of a closed convex set at a given point.
\begin{defn} 
\label{def:supportCone}
The support cone of a closed convex set  $\mathcal{K}$ at $x$ denoted by $\mathcal{L}_{\mathcal{K}}(x)$ is the smallest convex cone with vertex at the origin containing $\mathcal{K}-x$.
\end{defn}

\begin{lmma} 
\label{def:PartitionRn}
Let  $\mathcal{K}' $, $\mathcal{F}_J$, $\lbrace \gamma_i, i= 1,...,n \rbrace$ and  $\lbrace \beta_i, i= 1,...,n \rbrace$ defined as in Lemma \ref{def:Partitionface}. If $x$ is a point of $\mathcal{K}'$ belonging to the open face $F_J$, then:
\begin{itemize}
\item $\Pi^{-1}_{\mathcal{K}'}(x) = x + \mathcal{L}^o_{\mathcal{K}'}(x)= \Big \lbrace x + \sum_{i \notin J}c_i \gamma^i, c_i \geq 0\Big \rbrace = \Big \lbrace \sum_{j \in J}b_j \beta^j + \sum_{i \notin J}c_i \gamma^i, b_i > 0, c_i \geq 0\Big \rbrace$,

where $x=\sum_{j \in J} b_j\beta^j$.
 
\item The sets $\Pi^{-1}_{\mathcal{K}'}(x)$ are disjoint closed convex cones.

\item $\cup_{x \in \mathcal{K}'}\Pi^{-1}_{\mathcal{K}'}(x) = span(\mathcal{K})$
\end{itemize}
where $\mathcal{L}^o_{\mathcal{K}'}(x)$ denotes the dual of the support cone of $\mathcal{K}'$ at $x$.
\end{lmma}

Then any point in $span(\mathcal{K})$ projects onto an unique point in $\mathcal{K}'$ and belong to a unique non-negative orthant, or sector $S_J$
\begin{equation}
\label{eq:sector}
\mathcal{S}_J = \Big \lbrace  x \in \mathcal{R}^n: x =  \sum_{j \in J}b_j\beta^j + \sum_{j \notin J}c_j\gamma^j, b_j > 0, c_j \geq 0\Big \rbrace  
\end{equation}

Fig. \ref{fig:Partition} illustrates this result for $span(\mathcal{K})=\mathcal{R}^2$. Points in $S_J$ project  onto the subspace spanned by the vectors $\lbrace \beta^j, j\in J\rbrace$, that is on the face $F_J = \sum_{j \in J} b_j\beta^j$. Vectors belonging to $\mathcal{K}^o$ project onto the origin, vectors belonging to $\mathcal{K}$ project on themself, while each other vector of $\mathcal{R}^n$ project onto an unique face of  $\mathcal{K}$.

Therefore, if the sector $S_J$ containing the vector $u$ is known, then the projection of $u$ onto $\mathcal{K}$ can be easily computed as projection of $u$ onto the subspace spanned by the  $\lbrace \beta^j, j\in J\rbrace$. This reduces the problem of projecting $y$ onto $\mathcal{K}$ to the problem of finding the set of indices $\hat{J}$ such that the sector $S_{\hat{J}}$ contains $u$. 
The set complementary of $\hat{J}$ with respect to $\lbrace 1,...,m\rbrace$ corresponds to the indices of the constraints satisfied at equality in the optimal solution.

The algorithms described  in this section propose different strategies to find the optimal set $\hat{J}$.

\subsubsection{Early algorithms based on the properties of polyhedral convex cones}
The first algorithm addressing the problem of projecting a point $y \in \mathcal{R}^n$ onto a polyhedral convex cone $\mathcal{K} \subset \mathcal{R}^n$ by a non-asymptotic procedure dates back to work of  Wilhelmsen \cite{Wilhelmsen1976Nearest} in 1976. Wilhelmsen assume that the $m$ generators $\beta^i$ of the cone $\mathcal{K} = \Big \lbrace x \in \mathcal{R}^n: x = \sum_{i=1}^m b_i \beta^i, b_i >0 \Big \rbrace$  are known and propose an algorithm which compute a sequence of nearest points $x^k$ to $y$ in subcones $\mathcal{K}^k$ of $\mathcal{K}$. Each subcone $\mathcal{K}^k$ is chosen so that $x^k \in int(K^k)$ and is closer to $y$ than is $x^{k-1}$. This means that $x^k$ is in the near side of the supporting hyperplane of $\mathcal{K}^{k-1}$ with respect to $y$. The key step is to find $x^{k+1}$ given $x^k$ and the proposed procedure to do that is laborious.

Pshenichny and Danilin (1978) \cite{Pshenichny1978Numerical} proposed an algorithm similar to the one of Wilhelmsen which also converges in a finite number of steps. In both algorithm $m$ can be any integer even larger than $n$. A more efficient procedure, but with the more restrictive assumption that  $m\leq n$ has been proposed by Fraser and Massam in 1989.

\subsubsection{Mixed primal-dual basis algorithm}
\label{subsubsec:fraser}
Fraser and Massam \cite{Fraser1989Mixed} proposed an iterative algorithm to solve the general problem of projecting a data point $y \in \mathcal{R}^n$ onto a polyhedral convex cone $\mathcal{K} \subset \mathcal{R}^n$ generated by $m \leq n$ linear inequality constraints. 

Polyhedral convex cones generated by a number of inequalities at least equal to the dimension of the space they belong to have been the subject of section \ref{subsec:properties}. As seen there, the problem of projecting a data point onto this class of cones can be reduced to find the set of \textit{edges}, or generators of the cone, indexed by $\hat{J} \subseteq {1,...,n}$ such that the sector $S_{\hat{J}}$ contains the data point. 

To this goal, the set of edges of the polar cone $\lbrace \gamma^i, i=1,...,m \rbrace$ is completed by $n-m$  vectors orthogonal to $\lbrace \gamma^i, i=1,...,m \rbrace$ and orthonormal to each other. In the case of concave regression $m=n-2$, the set is completed by a constant function  $\gamma^{m+1}= \textbf{1}/||\textbf{1}||$, where $\textbf{1}$ is the $m$-dimensional unitary vector and by a linear function $\gamma^{m+2} = (x- \bar{x}\textbf{1})/||(x- \bar{x}\textbf{1})||$, where $x=(x_1,...,x_m)'$ and $\bar{x}= \sum_{i=1}^mx_i/m$. The set of vectors $\lbrace \gamma^i, i=1,...,n \rbrace$ form a basis for $\mathcal{R}^n$. 

Let the vectors $ \lbrace \beta_i, i=1,...,n \rbrace$ be the dual basis of the basis $\lbrace \gamma^i, i=1,...,n \rbrace$ as defined in (\ref{eq:dualbasis}). Fraser and Massam called the vectors $\beta^i$ and $\gamma^i$  primal and dual vectors respectively. A primal-dual basis for $\mathcal{R}^n$, associated to the set of indices $J \subseteq \lbrace 1,...,n\rbrace \equiv L$ is a basis $\mathcal{B}_{J} = [\alpha_1,...,\alpha_n]$ made up of a subset of the primal basis vectors $\lbrace  \beta_i \rbrace_{i \in J}$ and a complementary subset of the dual basis vector $\lbrace \gamma_i \rbrace_{i \in L\setminus J}$. For $n=m$ the primal basis vectors, corresponding to the edges of $\mathcal{K}$, are simply the columns of $-(A^T)^{-1}$.
Using the above definitions,  the problem of projecting a point $y \in \mathcal{R}^n$ onto the cone $\mathcal{K}$ can be formulated as follows.
\begin{thm}
The primal constrained quadratic minimization problem (\ref{eq:primalCQP}) is equivalent to the  problem of finding
\begin{equation} 
\label{eq:mpdb}
 \argmin_{ x \in \mathcal{K} } ||u-x||^2
\end{equation}
where $u = y - v$, with $v = \Pi(y|span(\gamma^{m+1},...,\gamma^{n}))$.
Denoting by $x_u$ the solution to this problem, the solution to the primal problem (\ref{eq:primalCQP}) is $\hat{x} = x_u + v$.
\end{thm}
 

Finding the sector containing $u$ is achieved moving along a fixed line joining an arbitrary  chosen initial point $x^0$ inside the cone or on its boundary to the data point $u$. 
By moving along a fixed line, many sectors are crossed: each time a sector is crossed the successive approximation $x^k$ is obtained by projecting the point on the line passing through $u$ and $x^0$ on the face $F_{J^k}$ of $\mathcal{K}$ (see Lemma \ref{def:Partitionface}) so that the distance  $||u-x^k||$ is decreasing in $k$. 

At each iteration each basis differs from another by one vector only and therefore the coordinates of $x^k$ onto the new basis are easy to calculate: what changes is that one coordinate becomes equal to zero and therefore there is not need of estimating the inverse of a matrix at each step. For this reason this algorithm is faster that the one of Wilhelmsen. 

Points belonging to the sector $S_{J^k}$ have non-negative coordinates in the mixed primal-dual basis $\mathcal{B}_{J^k}$ relative to the cone $\mathcal{K}$.  Therefore the procedure terminates when the coordinates of the data point in the primal dual basis $\mathcal{B}_{J^k}$ are all nonnegative, meaning that the point $x^k$ is on the face of the sector containing the data point $u$. 

The number of iterations needed is equal to the number of different sectors that the line joining the initial point to the data point has to cross. This number is bounded above by $2^n$. It is worth to remark that crossing a sector corresponds to a pivot step in the equivalent LCP. In fact, the net effect of a pivot step is  of moving from the point $x^k$ contained in the face $F_{J^k}$ of $\mathcal{K}$ to the point $x_{k+1}$ contained in the face $F_{J^{k+1}}$ of $\mathcal{K}$.




Ten years later,  Meyer \cite{Meyer1999Extension} generalized the algorithm of Fraser and Massam to the case of more constraints than dimensions, that is when $m>n$. 


\subsubsection{Critical index algorithm: Nearest point problem in simplicial cones} 
\label{subsec:critical}
Murty and Fathy (1982) \cite{Murty1982Critical} considered the general problem of projecting a given vector $y \in \mathcal{R}^n$ onto a \textit{simplicial cone} $\mathcal{K}\subset \mathcal{R}^n$. The definition of simplicial cone and \textit{pos cone}, on which the former definition is based are as follows.

\begin{defn}
\label{def:poscone}
The pos cone  generated by the vectors in $\Delta =  \lbrace  \delta^i, i = 1,...,m \rbrace$, denoted by $pos(\Delta)$, is the set $\lbrace x \in \mathcal{R}^n | x = \sum_{i=1}^m d_i \delta^i , d_i \geq  0\rbrace $.
\end{defn}

\begin{defn}
\label{def:simplicial}
A cone $\mathcal{K} \subset \mathcal{R}^n$ is said simplicial if it can be expressed as a positive linear span of $n$ linearly independent vectors $\Delta =  \lbrace  \delta^i, i = 1,...,n\rbrace$ in $\mathcal{R}^n$ (i.e., a basis for $\mathcal{R}^n$): $\mathcal{K} = pos(\Delta)$.
\end{defn}

For any point $x \in pos(\Delta)$ the vector $d = D^{-1}x$, where $D = [\delta^1,...,\delta^n]$ is called \textit{combination vector} corresponding to $x$. Therefore the projection $\hat{x}$ of $y$  onto $pos(\Delta)$ can be expressed as a nonnegative linear combination of the edges of the cone: $\hat{x} = \sum_{i=1}^n \hat{d}_i\delta^i$, where the optimal combination vector corresponding to  $\hat{x}$ is $\hat{d} = D^{-1}\hat{x}$. 

Murty and Fathy named the set of indices $\hat{J} \subseteq \lbrace 1,...,n \rbrace$ such that $\hat{d}_{i \in \hat{J}} > 0$ set of \textit{critical indices}.
Using the definitions of simplicial cone and combination vector, the original problem of projecting the point $y \in \mathcal{R}^n$ onto the cone $\mathcal{K}$ can be formulated as follows.
\begin{thm}
The primal constrained quadratic minimization problem (\ref{eq:primalCQP}) is equivalent to the  problem
\begin{equation} 
\label{eq:activeindex}
\hat d =  \argmin_{ d \geq 0 } (u-Dd)^T W (u-Dd) 
\end{equation}
where $u = \Pi(y|span(\mathcal{K}))$ and $D = [\gamma^1,...,\gamma^n]$, with $\lbrace \gamma^i=A_i^T, i = 1,...,m \rbrace$ and $\lbrace \gamma^i, i = m+1,...,n \rbrace$ defined as in section \ref{subsubsec:fraser}.

Denoting by $\hat{d}$ the solution to this problem, the solution to the primal problem (\ref{eq:primalCQP}) is $\hat{x} = y - D\hat{d} + v$.
\end{thm}
This formulation has the same structure of the dual formulation (\ref{eq:dualCQP}), where the combination vector $d$ in (\ref{eq:activeindex}) plays the same role as the dual variable $\lambda$ in (\ref{eq:dualCQP}). The only difference is that in (\ref{eq:dualCQP}) the matrix $A \in \mathcal{R}^{m\times n}$ is not squared as the matrix $D \in \mathcal{R}^{n\times n}$.  

As shown in section \ref{subsec:LCP} for the dual quadratic formulation (\ref{eq:dualCQP}), the formulation (\ref{eq:activeindex}) can be recasted as a LCP. This equivalency is important since the following theorem, proved in \cite{Murty1982Critical} applies to the LCP formulation of (\ref{eq:activeindex}).
\begin{thm}
\label{thm:reduction}
If a single critical index for the LCP problem of order $n$ is known, the problem can be reduced to a LCP of order $n-1$.
\end{thm}

The fact that finding a critical index reduces the dimension of the problem can be argued geometrically. Let $l$ be a critical index, then denoting by $NPP[\Gamma;u]$ the subproblem (\ref{eq:activeindex}), where $\Gamma = \lbrace \gamma^i,i=1,...,n\rbrace$ its solution is also the solution to the $NPP[\Gamma \cup \lbrace -\gamma^l\rbrace;u]$. Defining $\bar{u} = u - \frac{\gamma^l(u^T\gamma^l)}{||\gamma^l ||^2}$ and $\bar{\Gamma} = \lbrace \bar{\gamma}^1,...,\bar{\gamma}^{l-1}, \bar{\gamma}^{l+1},...,\bar{\gamma}^n \rbrace$, where $\bar{\gamma}^i = \gamma^i -  \frac{\gamma^l(u^T\gamma^l)}{||\gamma^l ||^2}\gamma^i$, than $\bar{\gamma}^i, i \in \lbrace 1,...,n \rbrace \setminus l$ is orthogonal to $\gamma^l$ and the cone $pos(\Gamma \cup \lbrace -\gamma^l\rbrace)$ is the direct sum of the full line generated by $\gamma^l$ and the simplicial cone $pos(\bar{\Gamma})$.

Solving $NPP[\bar{\Gamma},\bar{u}]$ is an $n-1$ dimensional problem. If $\hat{x}^*$ is the solution of the $NPP(\bar{\Gamma},\bar{u})$, then the solution  $\hat{x}$ to the $NPP[\Gamma;u]$ is obtained as $\hat{x} = \hat{x}^* + \frac{\gamma^l(u^T\gamma^l)}{||\gamma^l ||^2}$. 

By relying on Theorem \ref{thm:reduction}, the authors proposed an algorithm consisting of a subroutine to identify a critical index for the problem, followed by a subroutine which reduces the size of the problem once a critical index is found.
Since the solution $\hat{d}$ is the orthogonal projection onto the linear hull of $\lbrace \gamma^i, i \in \hat{J} \rbrace$, if $\hat{J}$ is known, the solution of the equivalent $LCP(q,M)$ and correspondingly the solution to $NPP[\Gamma;u]$ can be easily found.

The routine to identify a critical index operates on the $NPP[\Gamma;u]$ by exploiting the geometric properties of projection faces of a pos cone, whose definition is as follows.

\begin{defn}
\label{def:projectionFace} 
Let $S \subset \Gamma$. $pos(S)$ is a face of $pos(\Gamma)$ of dimension $|S|$. $pos(S)$ is said to be a projection face of $pos(\Gamma)$ if $\Pi(u|span(S)) \in pos(S)$.
\end{defn}

In the following we enunciated some theorems on which the critical index algorithm is based. Their proofs can be found in \cite{Murty1988Linear} (chapter 7).

\begin{thm}
\label{characterization_solution}
Let $S \subset \Gamma$, $S\neq 0$. The optimum solution of  $NPP[S;u]$ is in the relative interior of $pos(\Gamma)$if and only if the projection of $u$ onto the linear span of $S$ is in the relative interior of $pos(S)$: $\Pi(u|span(S)) \in  ri(pos(S))$.
\end{thm}

\begin{thm}
\label{projection_in_posCone}
Let $\hat{x} = D\hat{d}$ be the optimum solution of  $NPP[\Gamma;u]$. Let $\hat{J}$ the set of critical indices and $S = \lbrace \gamma^j, j \in \hat{J}\rbrace$. Then $pos(S)$ is a projection face of $pos(\Gamma)$.
\end{thm}

Theorem \ref{projection_in_posCone} tells that the projection onto the cone $\mathcal{K}^o$ belongs to the pos cone generated by the set $S$ of vectors  corresponding to critical indices and that such pos cone is a projection face. Therefore, is the set of critical index is known, for  Theorem \ref{characterization_solution} the solution can be computed as projection onto the linear subspace spanned by vectors in $S$.

The routine maintains a non empty subset of $\Gamma$ called the \textit{current set} denoted by $S$, and a point called the \textit{current point} denoted by $\bar{x}$.
At each stage of the routine $\bar{x} \in pos(S)$. When termination occurs the routine either finds the nearest point in $pos(\Gamma)$ to $u$ in which case the problem is completely solved or it finds a critical index of the problem. In the latter case an LCP of order $(n-1)$ can be constructed and the same routine can be applied to this smaller problem. Hence the unique solution of the original problem can be obtained after at most $n$ applications of the routine which finds a critical index.

A characterization useful to find a critical index or the solution to the problem is provided by the following theorem.
\begin{thm}
Let $\bar{x} \in pos(\Gamma)$ be such that $0 \in T(u,\bar{x})$, where $T(u,\bar{x})$ is the tangent hyperplane at $\bar{x}$ to the ball of center $u$ and radius $\bar{x}$. If there exists an index $j$ such that $(u-\bar{x})^T\gamma^i \leq 0 $ for all $i \neq j$ and $(u-\bar{x})^T\gamma^j$ then $j$ is a critical index of $NPP(\Gamma,u)$
\end{thm}
A characterization of the optimal solution in terms of separating hyperplanes is given by Robertson et al. \cite{Hardle1989Robertson}.
\begin{thm}
\label{thm:optimality}
A point $\bar{x} \in pos(\Gamma)$ is the nearest point in $pos(\Gamma)$to $y$ if and only if $0 \in T(y;\bar{x})$ and $(y-\bar{x})^T\gamma^j \leq 0$, $\forall j=1,...,n$, where $T(y,\hat{x})$ is the tangent hyperplane to $pos(\Gamma)$ in $\hat{x}$.
\end{thm}

The routine to find a critical index alternates distance reduction operations with line-search and projection steps to find a projection face. In practice, the routine starts by projecting on the closest edge to the data point. 
If the optimality condition is not satisfied, than the procedure iteratively adds vectors to $S$ and updates the point $\bar{x}$ while consistently reduces the distance between $u$ and  $\bar{x}$. The distance reduction operation is carried out efficiently by projecting onto the nonnegative hull of two vectors in $\mathcal{R}^n$, the current point $\bar{x}$ and a vector $\gamma^i$ satisfying one of the conditions given by the following theorem.

\begin{thm}
\label{thm:reduce_distance}
Given $\bar{x} \in pos(\Gamma)$, $\bar{x} \neq 0$ such that $0 \in T(y,\bar{x})$, if for some $i\in \lbrace 1,...,n\rbrace$ we have $(y-\bar{x})^T\gamma^i>0$ and either:
\begin{itemize}
\item $|| \bar{x} - y|| \leq ||\Pi(y|\gamma^i) -y ||$ and $\lbrace \bar{x},\gamma^i\rbrace$ is linearly independent, or
\item $y^T\gamma^i \leq 0$
\end{itemize}
then the projection of $y$ onto the linear hull of $\lbrace \bar{x},\gamma^i\rbrace$ is in the relative interior of $pos\lbrace \bar{x},\gamma^i\rbrace$
\end{thm}

Once such updates are not longer possibles, it employs a sequence of line-search steps and projections in the subspace spanned by the vectors in $S$ to find a projection face of the corresponding pos cone. This line-search  is in the same spirit than the one  proposed by Fraser and Massam since the goal is to reduce the distance  to the data point while keeping at the interior of a pos cone. In the particular case of concave regression, for which $m<n$, it can be implemented exactly in the same way.

This algorithm results to be much faster than the MPDB algorithm. The primary source of its computational efficiency  is in that it relies mostly on distance reduction operations and size reduction steps whose computational requirement is relatively small compared to the computational effort required to find a projection face through a line-search.


Recently, Liu and Fathy (2011) \cite{Liu2011Active} generalized the work of Murty and Fathy (1982) to polyhedral non-simplicial cones, hence allowing the set $\Gamma$ to contain more than $n$ vectors. What allows the generalization is the equivalence between the structure of the two problems through the concept of polar cone.  

The authors also proposed several strategies for efficient implementation mostly based on the mathematical properties of the entities involved. We have incorporated, where possible, these strategies, to all algorithm tested for objective evaluation of performances.

\subsubsection{Meyer's algorithm}
\label{subsubsec:hinge}
Meyer \cite{Meyer2013Simple} considered the general problem of projecting a data point $y \in \mathcal{R}^n$ onto a polyhedral  convex cone  $\mathcal{K} \subset \mathcal{R}^n$ generated by a finite number $m$  of linear inequalities. The problem is reduced to find the set of indices $\hat{J} \subset \lbrace 1,...,M \rbrace \equiv L $, where $M\leq m$ is the number of linearly independent constraints, corresponding to not saturated constraints at the solution. Meyer called these indices \textit{hinges}.

When $m\leq n$, the algorithm can be applied to both the primal and the dual formulation, whereas for $m>n$ it is applied to the dual formulation. In the following, since for the problem of concave regression  $m< n$ we consider how to solve the primal problem (\ref{eq:primalCQP}).

The search for  $\hat{J}$ is performed iteratively, starting with an initial guess by removing or adding one index at time, until the optimal solution is obtained. 
For each candidate $J^k$ two conditions are tested: the \textit{interior point condition} and \textit{optimality condition}.
 
The interior point  condition is satisfied when the current iterate belongs to the interior of $pos(S)$, that is when $x^k \in pos(S)$, where $S\subset \lbrace \beta^i, i \in J^k \rbrace$ . By using the following theorem
\begin{thm}
\label{thm:feasibility}
Let $S \subset \lbrace \beta^i,  i = 1...,m\rbrace$, $S \neq \varnothing$. The optimum solution of problem (\ref{eq:primalCQP}) is in the relative interior of $pos(S)$ if and only if $\Pi(y|span(S))$ is in the relative interior of $pos(S)$.
\end{thm}
$x^k$ can be computed as projection of $y$ onto the linear hull spanned by the vectors $\lbrace \beta^i, i \in J^k \rbrace$. 
If the feasibility condition  is not satisfied, the index $j\in L \setminus J^k$ corresponding to the most negative coefficient is added to $J^k$ and the interior point condition is checked again. 

Once the feasibility condition is satisfied, the optimality condition is tested by using the characterization given in Theorem \ref{thm:optimality}. If it is not satisfied, the vector $\beta^i, i \in J^k$ which most violates the condition is removed. The procedure continues until both conditions are satisfied. 

Convergence is guaranteed  by the fact that when the algorithm replaces just one edge, the Sum of the Squared Errors (SSE) after is less than the SSE before so that the algorithm never produces the same set of edges twice, which would result in an infinite loop.

In practice, each time an hinge is added, the best solution with $n+1$ hinges where the first $n$ hinges are already given is obtained. But this is not in general the best fit with $n+1$ hinges, so that some hinge may need to be changed. Therefore, the optimal solution can be interpreted as the best approximation with the biggest possible number of hinges. 

\section{Issues about effectivness for large-scale problems}

In this section, we discuss stenghts and limitations of the algorithms detailed above in solving large-scale instances of a particular kind of cone regression, the concave regression.  In particular, we consider computational issues related to numerical stability, computational cost and memory load as well as the suitability to take advantage of available good estimates and to be implemented in an online fashion.

\subsection{Suitability to  take advantage of available good estimates}

One general strategy for reducing the computational cost of a large-scale optimization problem is to use an initial guess, easier to calculate and close to the optimal solution.

Within the class of algorithms with asymptotic convergence, splitting-based methods work by activing each of the convex constraints repetitively and by combining them to obtain a sequence converging to a feasible point. Since the projection point $\Pi(y| \mathcal{K})$ is characterized by the variational inequality
\begin{equation}
\hat{x} = \Pi(y| \mathcal{K}) \in \mathcal{K}, \hspace{1cm} \forall x \in \mathcal{K}:  \langle y- \hat{x}, x-\hat{x}  \rangle  \leq 0
\end{equation}
the projection operator $\Pi( \cdot |\mathcal{K})$ is a closed contraction. Therefore the set of fixed points of $\Pi( \cdot |\mathcal{K})$ is exactly $\mathcal{K}$. 
This prevents the use of an initialization  point belonging to the feasible set as well as the use of multiscale strategies since there is not guarantee that the solution from a previous level does not belong to the feasible set. 

The same difficult arises when considering interior point algorithms, since them need to be initialized to an interior point. In \cite{Goldman1993Nonparametric},  Goldman  proved that the Dykstra's algorithm can potentially starts from better starting values than the given data point $y$. The author established the convergence to the nearest point to the primal cone from an arbitrary point in the intersection of $\mathcal{C}$ and the ball of radius $||y||$, where $\mathcal{C} = \lbrace x| x = y -A^T\lambda, \lambda \geq 0 \rbrace$, is a rotation of $\pi$ radiants of the dual cone $\mathcal{K}^o$ with its vertex translated at $y$. A point satisfying these conditions can be obtained efficiently by using distance reduction operations based on Theorem \ref{thm:reduce_distance}. It is worth to remark that this result can be easily interpreted in the active set framework. In fact, the Dykstra's algorithm can be undestood as a primal active set method and its solution is a primal dual feasible point. Therefore, any dual feasible point, that is every point belonging to the set $\mathcal{C}$, can be used as initialization.

All algorithm with time finite convergence detailed in the previous section are primal-dual and they reduce the problem of projecting a given data point onto a convex set to the problem of finding the set of indices corresponding to not saturated constraints at the solution. In general, they involve the selection of a subset from a collection of items, say $\hat{J} \subseteq \lbrace 1,...,m \rbrace$. With this formulation,  they  potentially allow to take advantage of a good estimate of the optimal active set. However, the adaptation is not rapid since the active set estimate is updated of one-by-one changes preventing this class of method from being effective general-purpose solvers for large-scale problems.

In the algorithm of Fraser and Massam, the set of successive approximations are obtained by moving along a fixed line connecting the initial guess to the data point. The number of iterations needed to reduce the data point reduces the number of sectors to be crossed to join the sector containing the data point. 

By contrast, in the algorithm of Meyer the proximity of the initial guess to the data point is not a good criterion selection for the initial guess. In fact, given an initial guess, the solution is attained by adding and/or removing one index at time until optimal solution is found. Taking into account that the optimal solution can be interpreted as the best approximation with the biggest possible number of hinges, if the optimal solution contains just a few hinges, than using the  empty set as an initial guess would result much faster than using the full set of possible hinges. On the contrary, if just a few constraints are satisfied at equality in the optimal solution, than the full set of indices will be a much better initial guess than the empty set. Therefore even if the choice of the initial guess may highly influence the performances, its choice depends on the data and there is not a well established criterion to fix it.

The Murty and Fathy's algorithm reduces the size of the problem each time a critical index is found. Therefore it is not compatible with strategies that take advantage of a good initial estimate since a good estimate does not lead to find a critical index faster.

To overcome the limitation of active set methods in taking advantage of a good estimate, Curtis et al. \cite{curtis2012} have recently proposed an euristic framework that allows for multiple simultaneous changes in the active-set estimate, which often leads to a rapid identification of the optimal set. However, there is not guarantee of the computational advantages for general problems and, furthermore, the authors recommend their approach when solving generic quadratic programming problems with many degrees of freedom, that is not the case of general concave regression problems.  



\subsubsection{PAV's inspired approximate solution}

To evaluate through numerical simulations the suitability to take advantage from good initial estimates, we propose an algorithm inspired to Pool Adjacent Violators (PAV), whose computational complexity is $\mathcal{O}$(n).
Starting from the original signal, violated constraints are removed one by one by projecting the current iterate onto the convex cone corresponding to the violated constraint until a primal feasible solution is obtained. Since the dual feasibility of each iterate is not guaranteed, the founded solution is not optimal. However, in our experience the solution is a very good approximation of the optimal solution.

\subsection{Suitability to be implemented in an online fashion}
Another strategy to deal with large projection problems would be to build and evaluate the solution incrementally according to the order of its input, as done in online methodologies developed for dynamic optimization \cite{Bhatia1996Dynamic} over a stream of input. Even if the input is given in advance, inputs are processed sequentially and the algorithm must respond in real-time with no knowledge of future input. Each new input may cause to rising or falling  constraints and the final solution is a sequence of feasible solutions, one for each time step, such that later solutions build on earlier solutions incrementally.

Of course this strategy requires that the algorithm respond in real-time for each new input, that would not possible when dealing with large matrix inverse computations. Let $\hat{x} \in \mathcal{R}^n$ be the projection of $y \in \mathcal{K}^n$ onto $\mathcal{K}$  and let $\hat{\bar{x}} \in \mathcal{K}^{n+1}$ be the projection of $\bar{y} \in \mathcal{R}^{n+1}$ onto $\mathcal{K}^{n+1}$. When a new element is added to the data point, a new constraint is added too, so that the constraint cone has a new edge. If this constraint corresponds to a critical index, that is to a constraint satisfied at equality in the optimal solution, then the projection face will be the same so that no further computing will be needed. On the contrary, if the new constraint does not correspond to a critical index, the projection face will change, including the edge corresponding to the new constraint and removing and adding some others. Therefore, the major difficulty faced by online strategy is the same faced in exploiting good estimates.

\subsection{Computational issues}
\label{subsec:computational_issues}
As highlithed in the previous section, despite the different strategies implemented by algorithms with time-finite convergence, generally an index at time is iteratively added or removed from the current set of indices $J^k$ until both the feasibility condition and the optimality condition are satisfied. Checking the optimality condition involves computing the inverse of a matrix that differs slightly from the matrix of the previous iteration. What "slightly" exactly means depends on the specific algorithm and it is detailed in the following.

The algorithm of Fraser and Massam involves the computation of a $n \times n$ fixed size matrix inverse at each iteration. The matrix to be inverted differs from the matrix of the previous iteration only for one column, being the change of the form $A \rightarrow (A + u \times v)$, where $u$ is an unitary vector with only one nonzero component corresponding to the index of the column to be changed, and  $v$ corresponds to the difference between the elements of the dual basis vector and the elements of the primal basis vector or viceversa. In this case the inverse can be updated efficiently by using the Sherman-Morrison formula: $(A + u \times v)^{-1} = \frac{z\times w}{1+\lambda}$, where $z= A^{-1}u$, $w=(A^{-1})^Tv$, $\lambda = v^Tz$. Therefore only two matrix multiplications and a scalar product are needed to compute the new inverse at each step.

The algorithm of Meyer, as well as the one of Liu and Fathi \cite{Liu2011Active} involve the computation of an inverse matrix of variable size at each iteration. The matrix to be inverted differs from the matrix of the previous iteration only for one column, which has been added or removed. Since the matrix to be inverted $A(J) \in \mathcal{R}^{r\times n}$, with $r\leq m$ is generally rectangular, the Moore-Penrose generalized inverse or pseudoinverse is computed:$A(J)^{\dag}= A(J)^T(A(J)A(J)^T)^{-1}$. 

In Matlab and Scilab, the computation of the pseudoinverse is based on the Singular Value Decomposition (SVD) and singular values lower than a tolerance are treated as zero. The advantage of this approach is that the pseudoinverse can be computed also for a nonsingular matrix. However, the method proposed by Liu and Fathi \cite{Liu2011Active} to improve the computational efficiency of their algorithm does not take advantage of SVD approach since it consists in updating the matrix $(A(J)A(J)^T)^{-1}$. If the matrix $A(J)$ is ill-conditioned, then the inverse cannot be computed with good accuracy and for the matrix $A(J)A^T(J)$ is even more so because this operation squares the condition number of the matrix $A(J)$.


A better solution would be to update directly the pseudoinverse. This can be achieved when a column is added by using the method proposed in \cite{Andelic:2006:KLM:1228509.1228513}
Let $A^T \in \mathcal{R}^{n\times r}$, $x \in \mathcal{R}^n$ and  $B = \begin{pmatrix}A^T&x\end{pmatrix} \in \mathcal{R}^{n\times (r+1) }$. The pseudoinverse of $B$ can be computed from the pseudoinverse of $A^T$ as follows.

$B^{\dag} = $
$\begin{pmatrix}
A^{\dag} - A^{\dag}xw^{\dag}\\
w^{\dag}
\end{pmatrix}$, where $w = (I-AA^{\dag})x$ and $w^{\dag} = \frac{w^T}{||w||^2}$.

Alternatively, the transformation 
$A(J)^T(A(J)A(J)^T)^{-1}y$ can be efficiently computed by a QR decomposition approach.
Let $A^T = \begin{pmatrix} Q_{11}  Q_{12} \end{pmatrix}\begin{pmatrix}R_{11} \\ 0 \end{pmatrix}$ be the QR decomposition of $A^T$, where $R_{11}$ is an $m\times m$ invertible upper triangular matrix. Then: $A(J)^T(A(J)A(J)^T)^{-1}=Q_{11}(R^T_11)^{-1}$. The inverse of an upper triangular matrix can be efficiently implemented by a left matrix division or by more sophisticated methods as the one proposed in \cite{mahfoudhi2012fast}.

Courrieu \cite{Courrieu2008Fast} proposed a method based  on the full rank Cholesky decomposition which has a computation time substantially shorter of the method based on SVD decomposition.
The two main operations on which his method is based are the full rank Cholesky factorization of $A^TA$ and the inversion of $L^TL$, where $L$ is a full rank matrix .
On a serial processor these computations are of complexity order $\mathcal{O}(n^3)$ and $\mathcal{O}(m^3)$ respectively. However, in a parallel architecture, with as many processor as necessary, the time complexity for Cholesky factorization of $A^TA$ could reduce to $\mathcal{O}(n)$, while the time complexity for the inversion of the symmetric positive definite matrix $L^TL$ could reduce to $\mathcal{O}(log(r))$. However, the computational advantage of this method can be appreciated only when $r<<n$, since the inverse of a $r \times r$ matrix has to be computed, which is not in general the case, specially for concave regression problems.

The method proposed by Zhu and Li \cite{Zhu2007Recursive} for recursive constrained least squares problems, found that the exact solution of Linear Equality-constrained Least Squares can be obtained by the same recursion as for the unconstrained problem, provided that the Rescricted Least Squares procedure is appropriately initialized. However, even this approach offer significant advantages in term of computational cost only when the number of constraints $m$ is small, which is not the case for large-scale concave regression problems.



\section{Improving the active set framework for concave regression problems}
An active set algorithm for solving the concave regression problem generates a sequence of quasi stationary points. A primal (dual) feasible active set algorithm generates a sequence of primal (dual) feasible quasi stationary points with decreasing objective function values and terminates when the dual (primal) feasibility is satisfied.
An active set $J$ induces a unique partition of the indices $\lbrace1,...,n\rbrace$ into blocks. A block $B$ of such partition is a set of consecutive integers, say, $\lbrace p,p+1,...,q \rbrace$, such that the index $i$ of the constraint $x_{i+1} - x_i \leq x_{i+2} - x_{i+1}$ is in $J$ for each $i$ such that $p \leq i \leq q-2$. Conversly any such partition of indices determines a unique active set.
Denoting by $\lambda_i$  the multiplier associated with the ith constraint, the Karush-Kuhn-Tucker can be written as:

\vspace{0.5cm}

$\left\lbrace
\begin{array}{llllllcl}
x_{i+1} - x_i \leq x_{i+2} - x_{i+1} \quad i = 1,...,n-2\\
y_1 - x_1 = \lambda_1\\
y_2 - x_2 = - 2\lambda_1 + \lambda_2 \\
y_3 - x_3 = \lambda_1 - 2\lambda_2 + \lambda_3 \\

y_{n-1} - x_{n-1} = \lambda_{n-4} - 2\lambda_{n-3} + \lambda_{n-2}\\
y_{n-1} - x_{n-1} = \lambda_{n-3} - 2\lambda_{n-2} \\
y_n - x_n = \lambda_{n-2} \\
\lambda_i \geq 0 \quad \quad  i = 1,...,n-2\\
\lambda_i(2x_{i+1} - x_i - x_{i+2}) = 0 \quad i = 1,...,n-2\\
\end{array}\right.$ 

\vspace{0.3cm}

It is easy to show that: $\sum_{i=1}^n i(y_i-x_i) = 0$ and $\sum_{i=1}^n (y_i-x_i) = 0$. Knowing that each block can be represented by an affine function $x_i = \alpha + \beta i$, the case  of  blocks the above systems can be written as:
 
\vspace{0.5cm}

$\left\lbrace
\begin{array}{lllllllcl}
\sum_{i=1}^{n_1}y_i = \alpha^1 \sum_{i=1}^{n_1}i + \beta^1 n_1 + \lambda_{n_1} \\

\sum_{i={n_1+1}}^{n}y_i = \alpha^2 \sum_{i={n_1+1}}^{n}(i-n_1) + \beta^2 (n-n_1) - \lambda_{n1} \\

\sum_{i={n_1+1}}^{n}y_i = \alpha^2 \sum_{i={n_1+1}}^{n}(i-n_1) + \beta^2 (n-n_1) - \lambda_{n1} \\

\sum_{i={n_1+1}}^{n}(i-n_i)y_i = \alpha^2 \sum_{i={n_1+1}}^{n}(i-n_1)^2 +  \beta^2 \sum_{i={n_1+1}^{n}} (i-n_1) - \lambda_{n1} \\

\lambda^1n_1 + \beta^1-\alpha^2-\beta^2=0
\end{array}\right.$ 

\vspace{0.3cm}

Therefore, for each block the unknown variables to be computed are $\alpha, \beta, \lambda$. 
The systems to be solved can be written as $A\boldsymbol{x} = \boldsymbol{b}$, where $\boldsymbol{x} = (\alpha^1,\beta^1,\alpha^2,\beta^2,\lambda_{n_1})$, $\boldsymbol{b} = (\sum_{i=1}^{n_1}y_i,\sum_{i={n_1+1}}^{n}y_i,\sum_{i={n_1+1}}^ny_i,\sum_{i={n_1+1}}^n(i-n_i)y_i , 0)$ and 

\vspace{0.5cm}

$A = $$\begin{pmatrix}
       \sum_{i=1}^{n_1} i & n_1 & 0           & 0     				 &  1  \\[0.3em]
       0 & 0           &    	\sum_{i=1}^{n-n_1} i & n-n_1 	  		 & -1  \\[0.3em]
        \sum_{i=1}^{n_1} i^2          &  \sum_{i=1}^{n_1} i & 0    & 0				 & n_1 \\[0.3em]
       0           & 0 &\sum_{i=1}^{n_2-n_1} i^2 & \sum_{i=1}^{n_2-n_1} i    &  0  \\[0.3em]
       n_1+1       & 1 & -1 & -1    						 &  0  \\[0.3em]
     \end{pmatrix}$
 
\vspace{0.3cm}

In general, if $k$ is the number of blocks, than the system to be solved has size $3k-1$.
As observed in \cite{Kuosmanen2008Representation}, in the case of concave regression the number of blocks at the solution is usually much lower than $n$. Therefore, an active set algorithm that start with an empty active set, should found the solution without the need of inverting large matrices.

\section{Comparative results}
\label{sec:experiments}

In Tab. \ref{tab:1} we compare qualitatively the algorithms analyzed above in terms of their formulation (primal, dual, or primal-dual), their possibility  to be initialized and their major limitations in dealing with large scale problems.  All analyzed methods are dual or primal-dual: dual methods cannot be initialized, whereas initialization in primal-dual methods is allowed but constrained. The major limitation of asymptotic convergence methods when dealing with large problems is the convergence, whereas time-finite convergence methods become too slow and numerical instable because of accumulation errors. In the following, we provide evidence of these limitations through numerical simulations reported in the Appendix   and we compare quantitatively their performances  in term of distance from the solution for one second, one minute and ten minutes of CPU time.
Instead of using only unstructured random data as data-test, we considered signals of increasing difficulty level varying from a noised concave signal to a concave/convex noised signal.
More precisely, we considered three signals, whose equations and plots are given in Fig. \ref{signals} of the Appendix and, for each of them, we considered three different sizes: $n \in \lbrace 50, 500, 1000 \rbrace$
and three increasing levels of noise (standard deviations $\sigma \in \lbrace 0.01, 0.1, 0.5 \rbrace$). To evaluate robustness against initialization variation for time-finite active set methods, we considerer three different initializations  of the active set $J$: the empty set, the set of $m$ indexes corresponding to the linear inequality constraints, the set of not saturated constraints obtained by using the algorithm inspired to PAV described in the previous section.  In the class of algorithms with asymptotic convergence, the most efficient results to be ADMM. This is evident even for very small size data, when using difficult signal such as near-convex signals or very noised signals (see Fig. \ref{size50L2S2} 
and Fig. \ref{size50L2S3} 
of the Appendix). For noised concave signals (Fig. \ref{size50L2S1} 
of the Appendix), the computational efficiency of ADMM is more evident in presence of an high level of noise. Already for signals of size $500$ the performance of ADMM are not very good: convergence ($SSE<10^{-6}$) is not completely attained.
The algorithm of Meyer is very sensitive to the initialization. It gives good performances when the signal is a Gaussian white noise and the initial active set is empty since most of constraints are expected to be saturated at the solution (see Fig. \ref{size50L2S3} 
and Fig. \ref{size500L2S3} of the Appendix). 
Given a good initialization, the MPDB algorithm allows to compute the exact solution faster than other methods. However, this algorithm is not numerically stable since it require to compute  the inverse of a matrix at each iteration. As explained in section \ref{subsec:computational_issues}, this is than incrementally by using the Shermann-Morrison formula. However, numerical round-off errors cumulate so that, in our implementation, the exact inverse is computed each $150$ iterations. As it can be observed in Fig. \ref{size1000L2S1}
, Fig. \ref{size1000L2S2} 
and Fig. \ref{size1000L2S3} 
of the Appendix, that refer to signals of size $1000$,  sometimes the round-off error dominates the calculation and the distance to the solution increases instead of decreasing. 
These results demonstrated that, although the theoretical and practical advances in recent years, the use of shape-constrained nonparametric techniques for solving large scale problems (more than many thousand of points) is still limited by computational issues. To deal with very large scale problem,  up to a million of points, matrix inverse calculation should be avoided and more efforts  should be devoted to find a way to better initialized primal-dual methods, by computing approximate solutions and exploiting multi-scale strategies.

\par

\section{Conclusions}
\label{sec:conclusions}

In this paper we have stated, analyzed and compared qualitatively and quantitatively several optimization approaches for solving the cone regression problem. We have distinguished two broad classes of methods. On one side, methods with asymptotic convergence that rest on the properties of proximity operators and act by finding the solution as the limit of a sequence of successive approximations. On the other side, methods with finite-time convergence that exploit the geometric properties of polyhedral convex cones and find the exact solution as non-negative linear combination of functions, forming a basis in a specified finite dimensional space. 
Simulations up to one thousand of points have demonstrated that the choice of the optimization approach severely impact algorithmic performances. In particular, it has emerged that methods with time-finite convergence are much more efficient with respect to asymptotic-convergence methods.  However, from this study it emerged that they face a twofold difficulty to cope with large-scale optimization: the first difficulty arises from the fact that all algorithm of this class modify the active set estimate one-by-one, making the adaptation of the initial active set estimation very slow; the second difficulty lies in the fact they involve the computation of the inverse or the pseudoinverse of a matrix at each variation of the active set. Although there exists many methods to do that efficiently when the matrix rank is much lower that the size of the data, this condition cannot be assured in general. Incremental strategies to reduce the cost of computing the inverse of a matrix when the inverse of a slightly different matrix is known, are bounded by round-off error accumulation in an iterative setting.  The results of this study suggest that to be able to trait very large-scale problems (up to a million of points) further research should focus on finding a way to exploit classical multi-scale strategies and to compute an approximate solution through a penalization or splitting method without involving any matrix inverse calculation.
\begin{table*}
\caption{Qualitative comparison of all algorithms}
\label{tab:1}       
\begin{tabular}{llllll}
\hline\noalign{\smallskip}
Algorithm & Type & Formulation & Initialization & Limitation \\
\noalign{\smallskip}\hline\noalign{\smallskip}

Hildret & dual & \ref{eq:dualCQP} & not possible& convergence\\
Dykstra & primal-dual & \ref{eq:primalCQP} &  constrained & convergence \\
ADMM & dual & \ref{eq:admm}  & not possible& convergence\\
LSPS & primal & \ref{eq:lsps} &  constrained & convergence \\
Uzawa & dual & \ref{eq:uzawa} & not possible& convergence \\
MPDB & primal-dual & \ref{eq:mpdb} &  constrained & slow or instable \\
Meyer & primal-dual & \ref{eq:primalCQP} & constrained and not robust & slow or instable \\
Active Index & dual & \ref{eq:activeindex} & not possible& slow or instable \\
\noalign{\smallskip}\hline
\end{tabular}

\end{table*}
\par

\clearpage
\newpage 

\section{Appendix*}
\subsubsection{Signals used for comparative evaluations}

$S_1(z) = $$\left\lbrace
\begin{array}{lll}
 2nsin(\frac{24}{5n}z) \quad \quad z = 1,...,\frac{n}{3}\\
\alpha + \beta z \quad \quad z = \frac{n}{3}+1,...,\frac{2n}{3} \\
 \gamma z^3 +  \delta \quad  \quad z = \frac{2n}{3}+1,...,n
\end{array}\right.$ 
\vspace{0.1cm}
where $\beta = \frac{1}{10}$, $\alpha = 2n sin(\frac{8}{5}) - \beta n$, $\gamma = \frac{-2}{n^2}$ and $\delta = \alpha + \beta \frac{2n}{3} - \gamma \frac{8n^3}{27}$.

\vspace{0.3cm}

$S_2(x) = \mathcal{N}(\mu,\,\sigma^2)$

\vspace{0.3cm}
$S_3(z) =  sinc(\frac{6}{n}z-1)$
\begin{figure}[!h]
\begin{tabular}{llllcc}
\includegraphics[width=48mm]{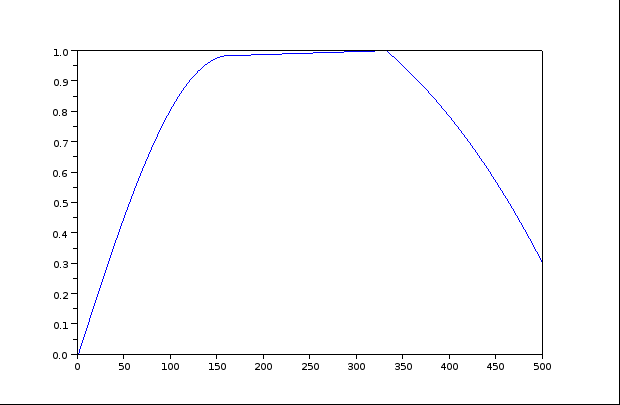}&
\includegraphics[width=48mm]{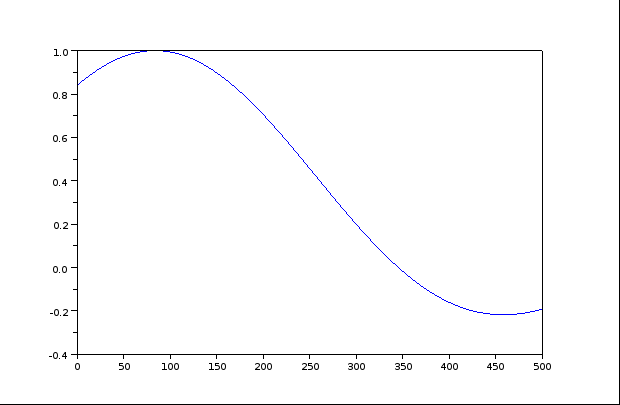}\\
\includegraphics[width=48mm]{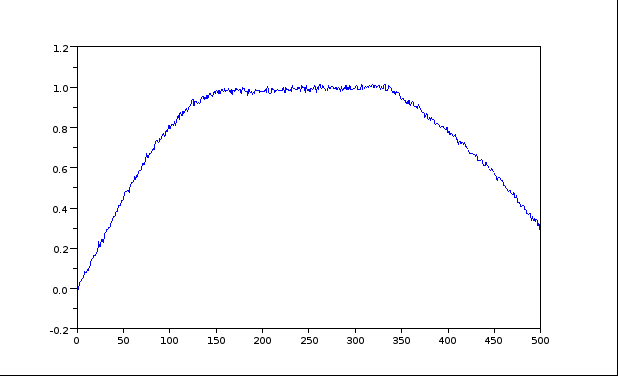}&
\includegraphics[width=48mm]{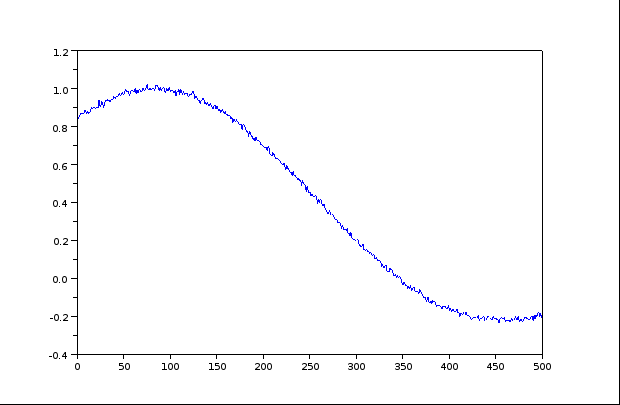}\\
\includegraphics[width=48mm]{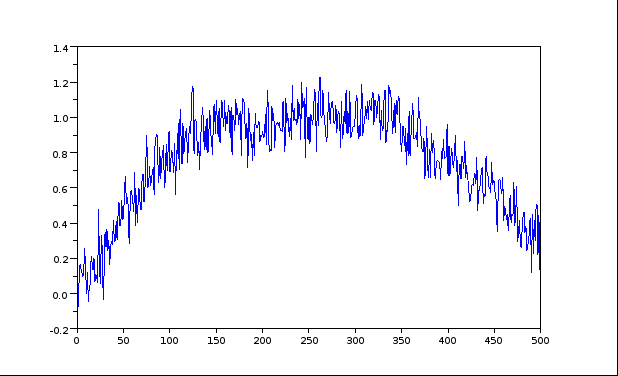}&
\includegraphics[width=48mm]{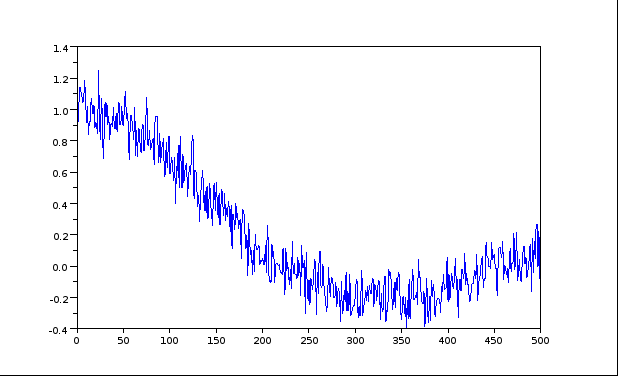}\\
\includegraphics[width=48mm]{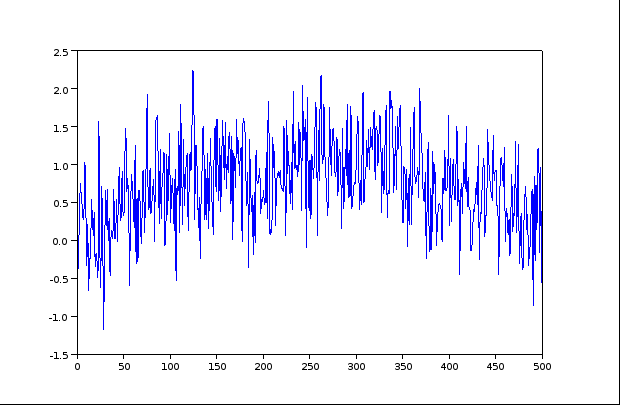}&
\includegraphics[width=48mm]{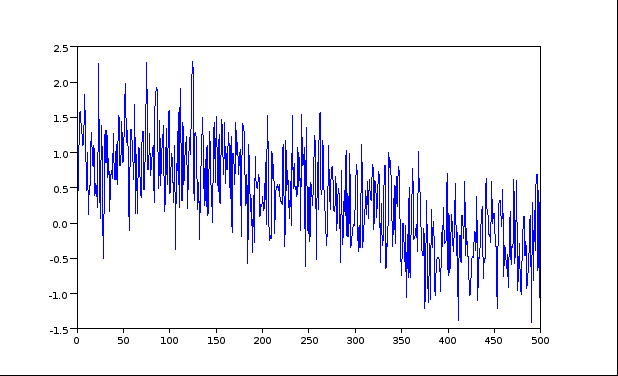}
\end{tabular}
\caption{Signals $S_1$ (top left) and $S_2$ (top rigth) to which has been added white noise with three different values of standard deviation $\sigma = 0.01$, $\sigma = 0.1$, and $\sigma = 0.5$.}
\label{signals}
\end{figure}

\clearpage
\newpage

\subsubsection{Comparative evaluations on signals of size 50}

\begin{figure*}[!ht]
\begin{tabular}{lllcc}
\includegraphics[width=70mm]{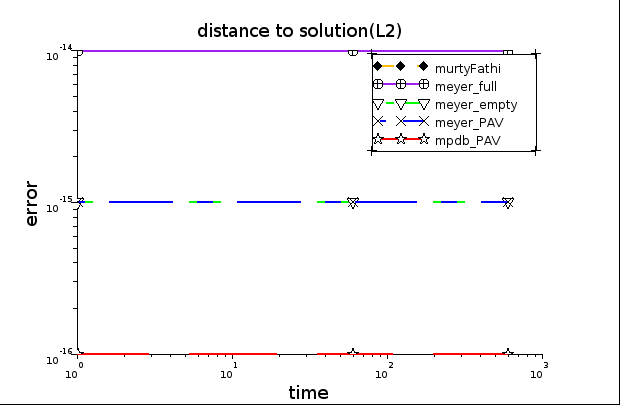}&
\includegraphics[width=70mm]{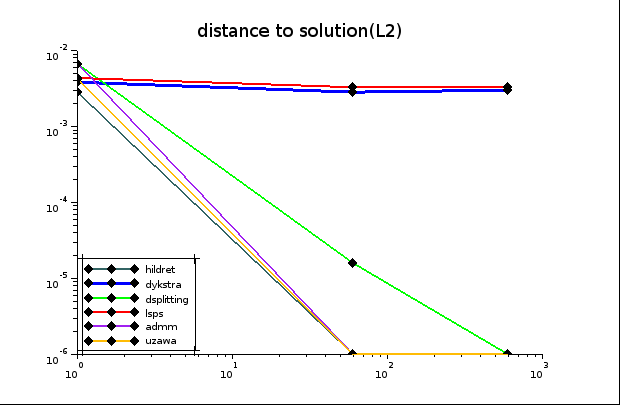}\\
\includegraphics[width=70mm]{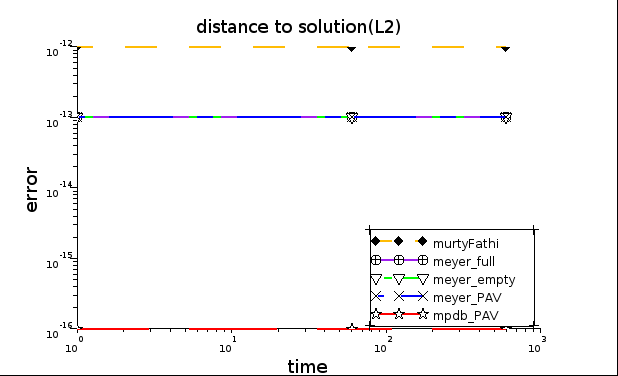}&
\includegraphics[width=70mm]{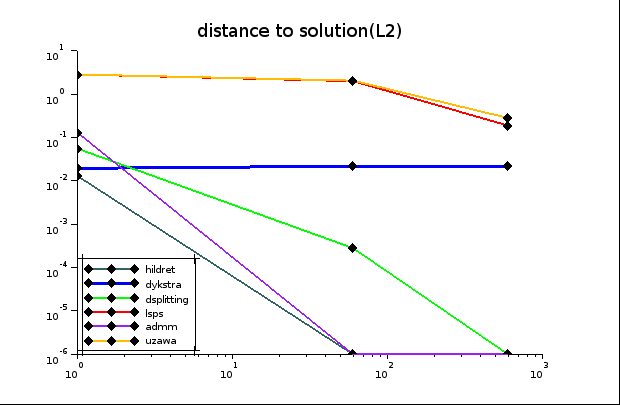}\\
\includegraphics[width=70mm]{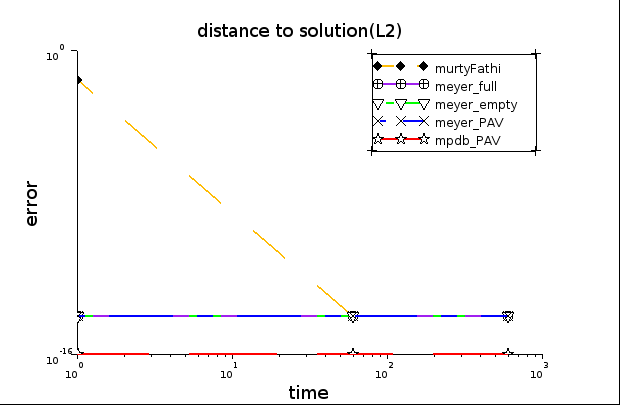}&
\includegraphics[width=70mm]{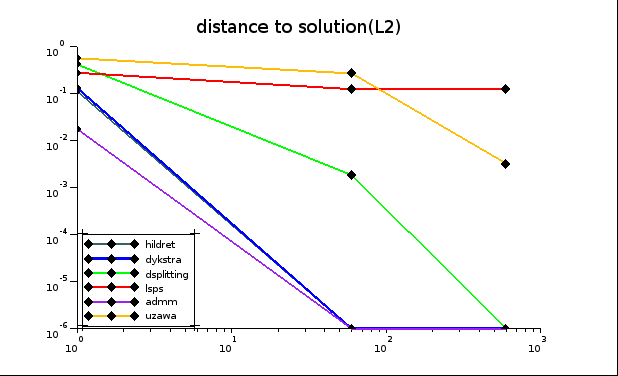}
\end{tabular}
\caption{Distance to the solution ($L_2$ norm) for a signal of type $S1$ of size $50$. From up to the bottom, three increasing level of noise have been added. (Left) Algorithms with time-finite convergence: all them converge istantanely. (Right) Algorithms with asymptotic convergence: ADMM is more robust to noise. LSPS and Dykstra use a parametrization in the primal parameter space which causes numerical round-off errors cumulate so that the fitted values does not satisfy the constraints of the dual problem and convergence is not attained.}
\label{size50L2S1}
\end{figure*}

\newpage 

\begin{figure*}
\begin{tabular}{llllcc}
\includegraphics[width=70mm]{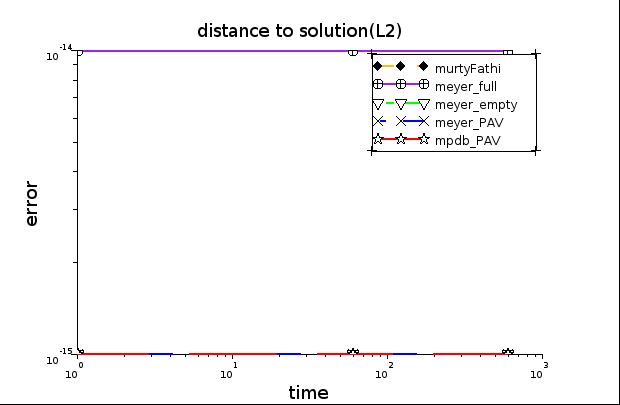}&
\includegraphics[width=70mm]{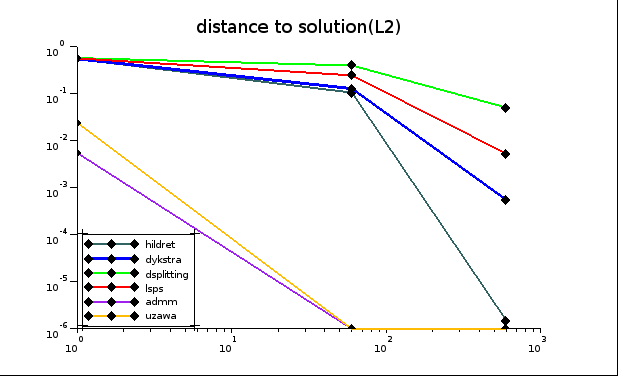}\\
\includegraphics[width=70mm]{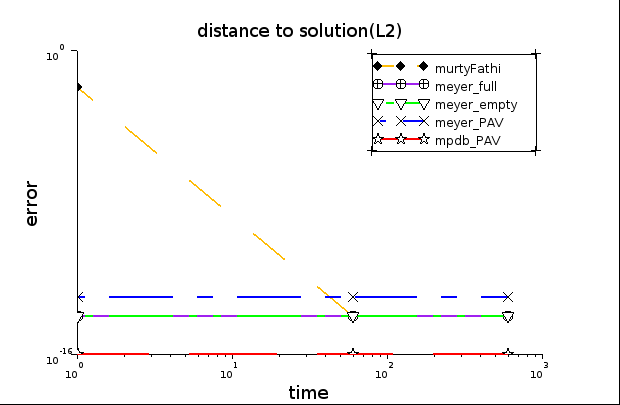}&
\includegraphics[width=70mm]{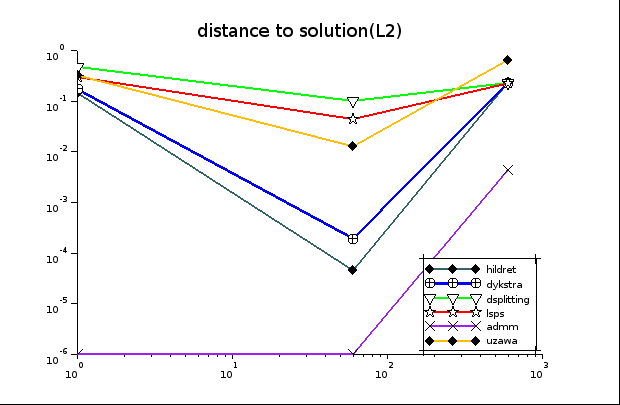}\\
\includegraphics[width=70mm]{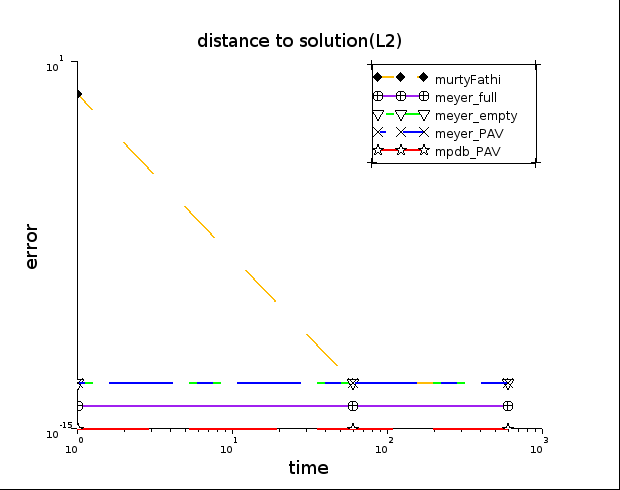}&
\includegraphics[width=70mm]{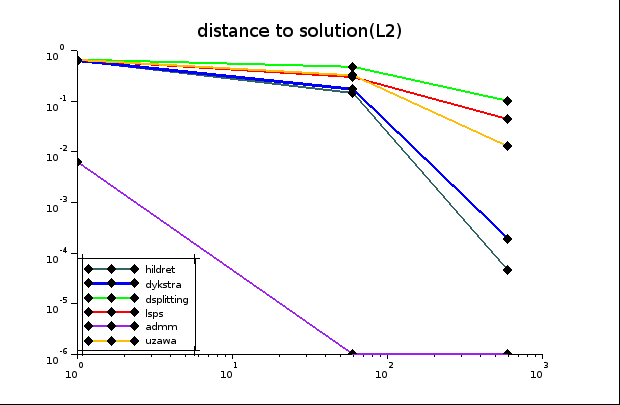}
\end{tabular}
\caption{Distance to the solution ($L_2$ norm) for a signal of type  $S_2$ of size $50$. From up to the bottom, three increasing level of noise have been added. (Left) Algorithms with time-finite convergence: all them converge istantanely. (Right) Algorithms with asymptotic convergence: both ADMM and Hildret attain the solution when not noise is added but ADMM is much more robust to noise.}
\label{size50L2S2}
\end{figure*}

\newpage 

\begin{figure*}
\begin{tabular}{llllcc}
\includegraphics[width=70mm]{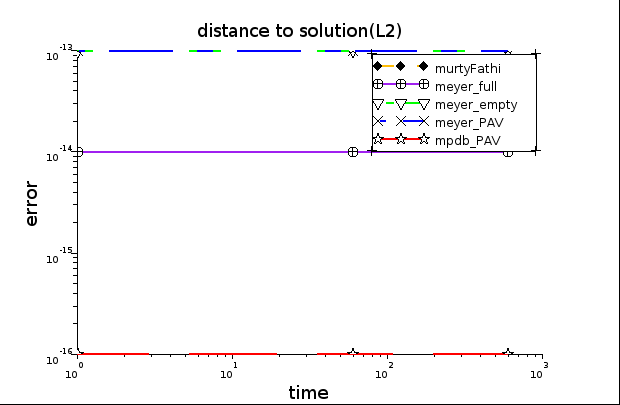}&
\includegraphics[width=70mm]{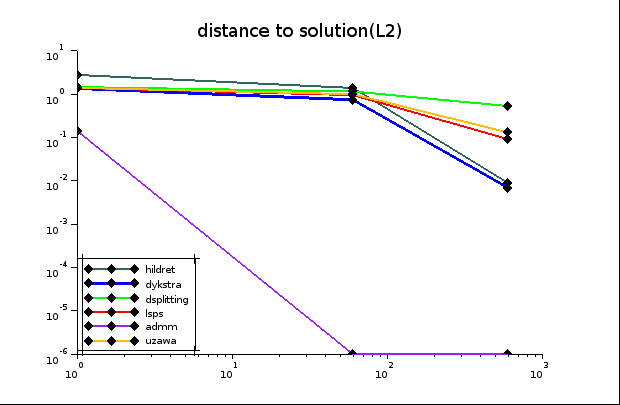}\\
\includegraphics[width=70mm]{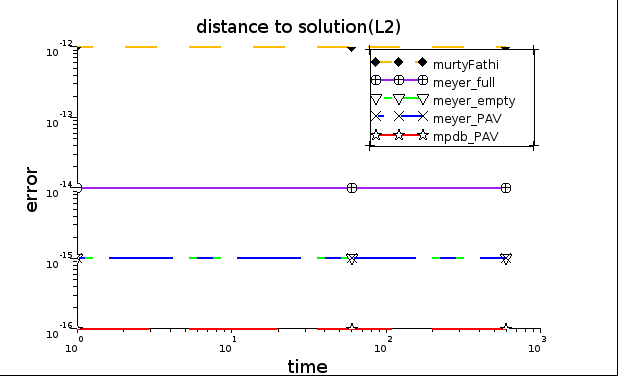}&
\includegraphics[width=70mm]{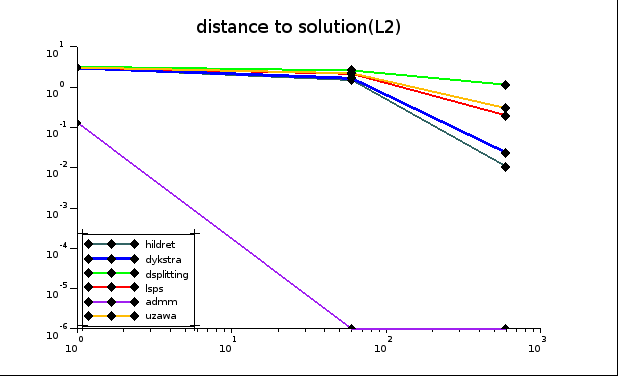}\\
\includegraphics[width=70mm]{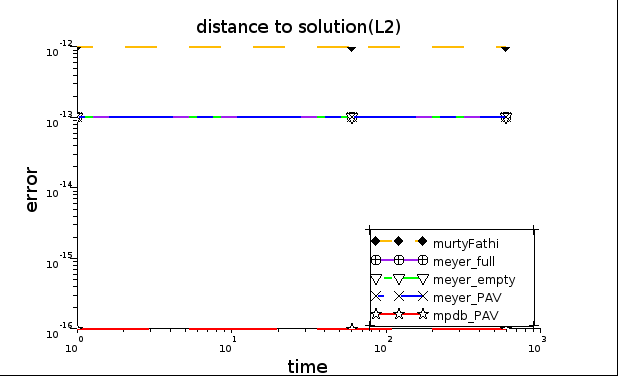}&
\includegraphics[width=70mm]{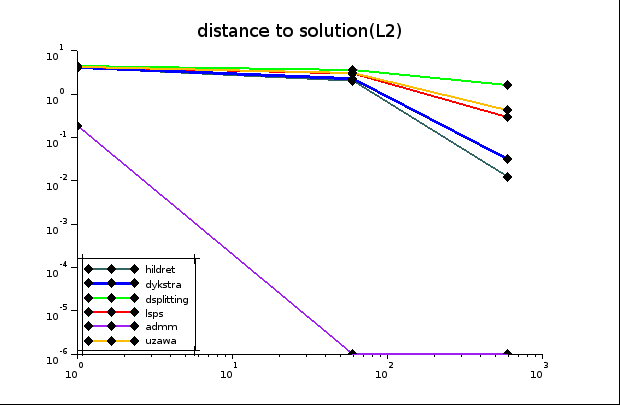}
\end{tabular}
\caption{Distance to the solution ($L_2$ norm) for Gaussian white noise signals of size $50$. (Left) Algorithms with time-finite convergence: all them converge istantanely. (Right) Algorithms with asymptotic convergence: ADMM is the only algorithm that converges.}
\label{size50L2S3}
\end{figure*}

\newpage 
\onecolumn
\subsubsection{Comparative evaluations on signals of  size 500}
\begin{figure*}[!ht]
\begin{tabular}{lllcc}
\includegraphics[width=70mm]{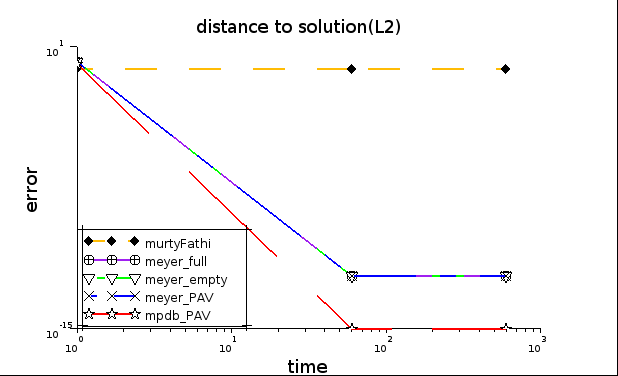}&
\includegraphics[width=70mm]{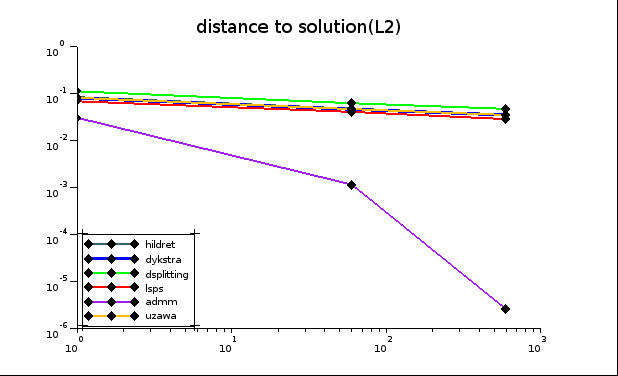}\\
\includegraphics[width=70mm]{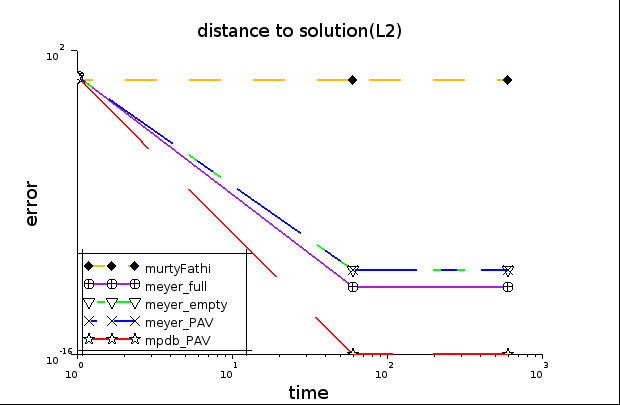}&
\includegraphics[width=70mm]{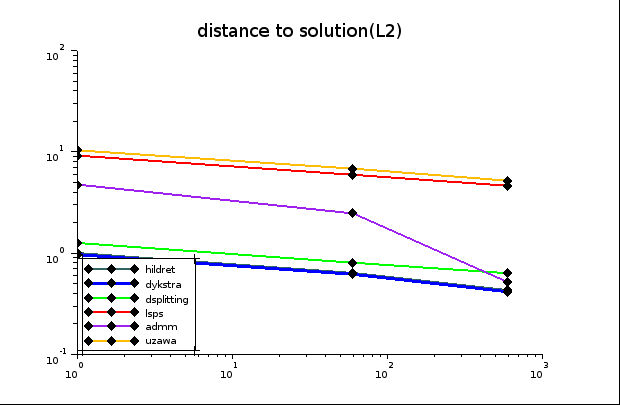}\\
\includegraphics[width=70mm]{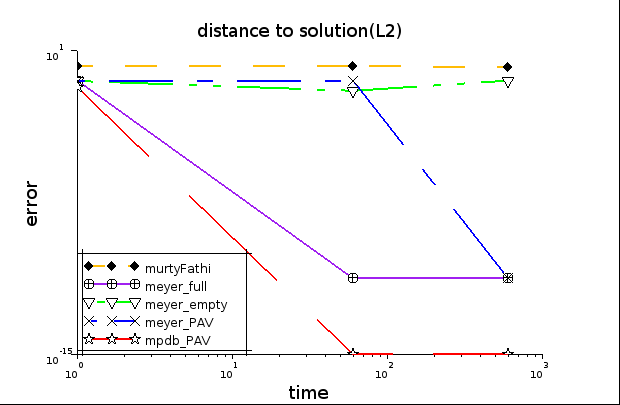}&
\includegraphics[width=70mm]{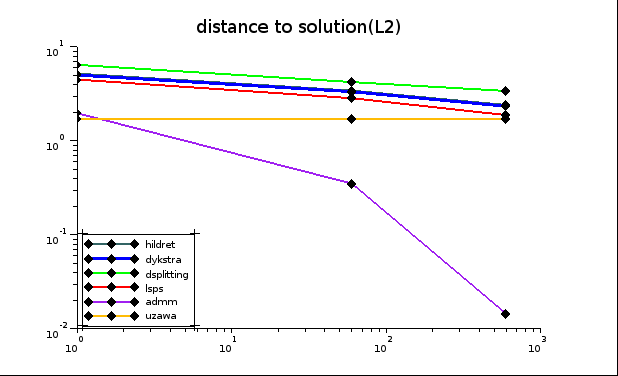}
\end{tabular}
\caption{Distance to the solution ($L_2$ norm) for a signal of type $S_1$ of size $500$. From up to the bottom, three increasing level of noise have been added. (Left) Algorithms with time-finite convergence:  Meyer's algorithm is very sensitive to the initialization. The best performance are achieved by MPDB with a PAV approximate initialization. (Right) Algorithms with asymptotic convergence: ADMM converges only for when the signal is slighly noised.}
\label{size500L2S1}
\end{figure*}

\newpage 

\begin{figure*}
\begin{tabular}{llllcc}
\includegraphics[width=70mm]{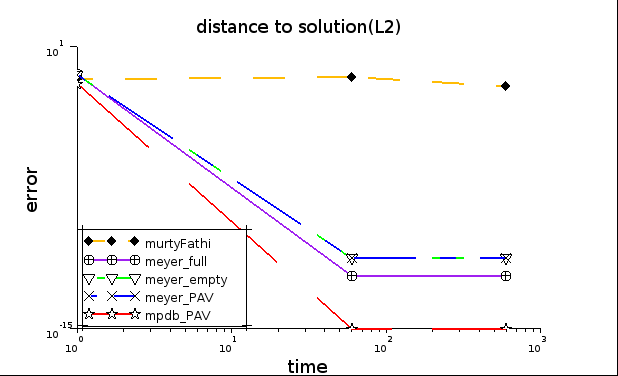}&
\includegraphics[width=70mm]{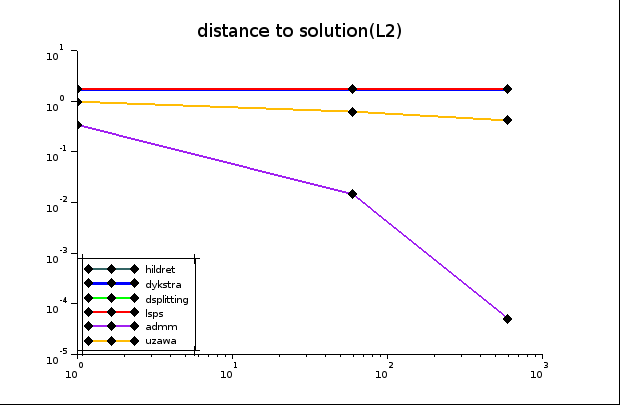}\\
\includegraphics[width=70mm]{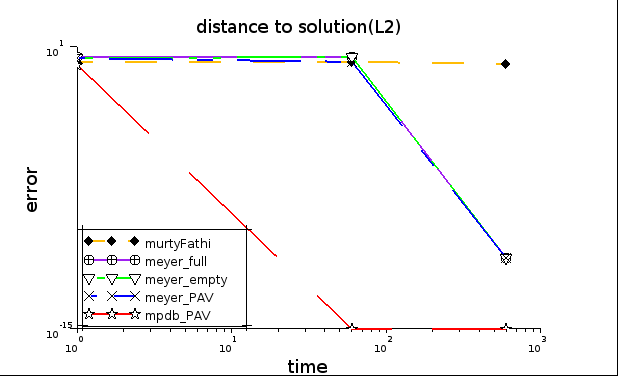}&
\includegraphics[width=70mm]{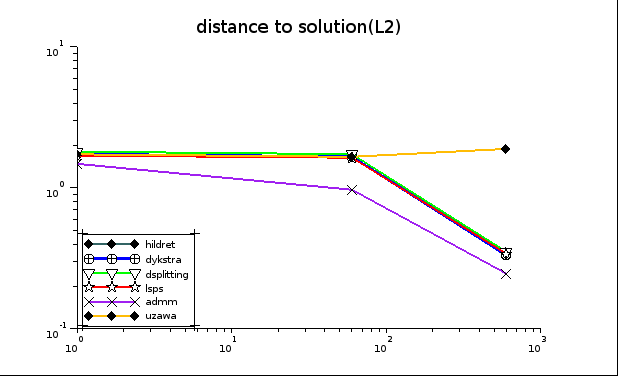}\\
\includegraphics[width=70mm]{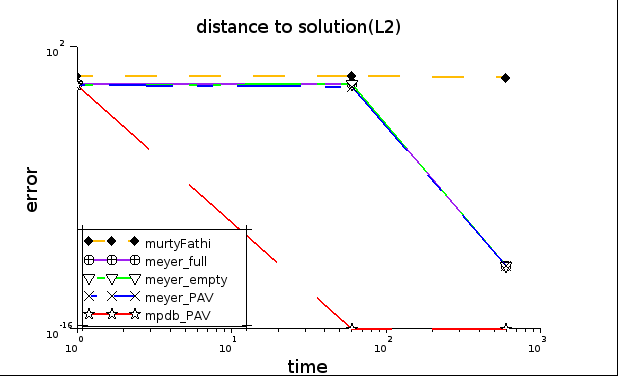}&
\includegraphics[width=70mm]{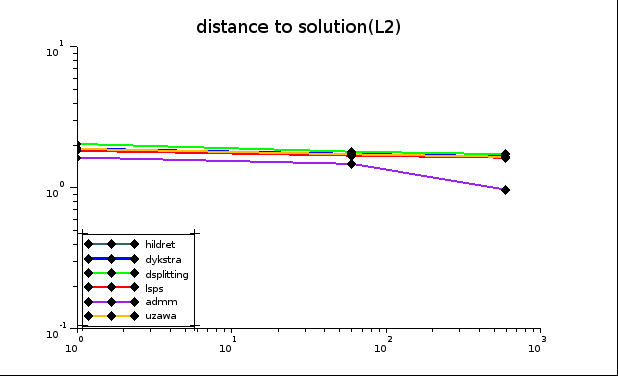}
\end{tabular}
\caption{Distance to the solution ($L_2$ norm) for a signal of type  $S_2$ of size $500$. From up to the bottom, three increasing level of noise have been added. (Left) Algorithms with time-finite convergence: the best performance are achieved by MPDB with a PAV's inspired initialization. (Right) Algorithms with asymptotic convergence: no algorithm converges.}
\label{size500L2S2}
\end{figure*}

\newpage 

\begin{figure*}
\begin{tabular}{llllcc}
\includegraphics[width=70mm]{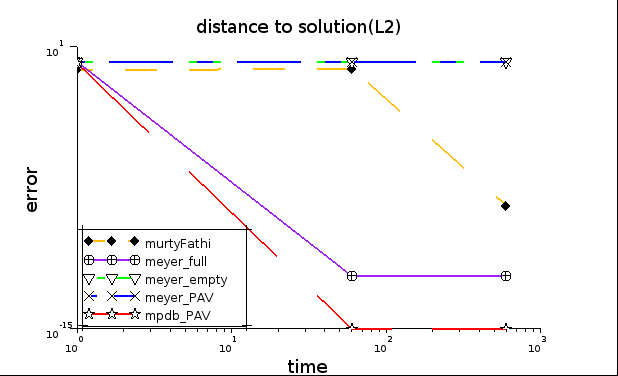}&
\includegraphics[width=70mm]{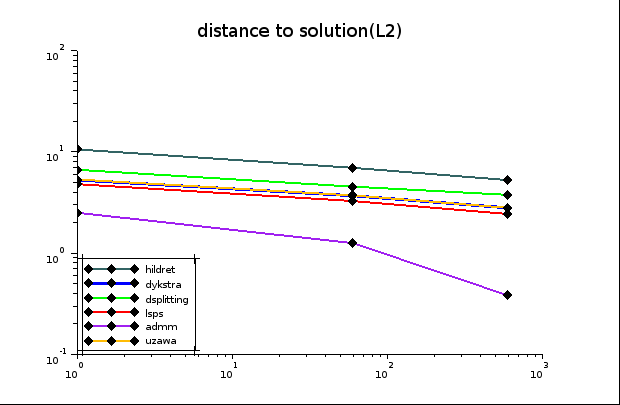}\\
\includegraphics[width=70mm]{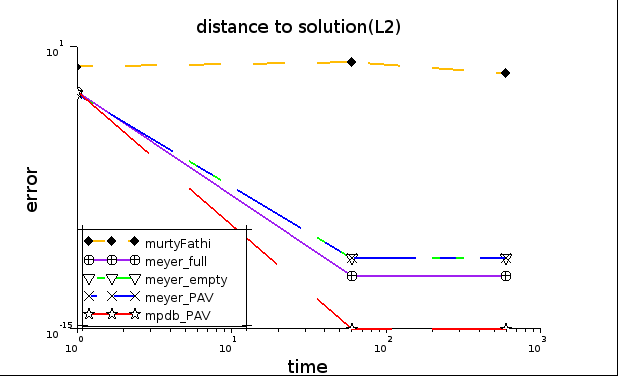}&
\includegraphics[width=70mm]{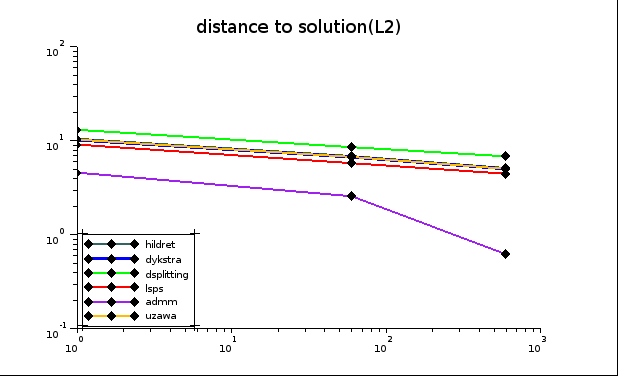}\\
\includegraphics[width=70mm]{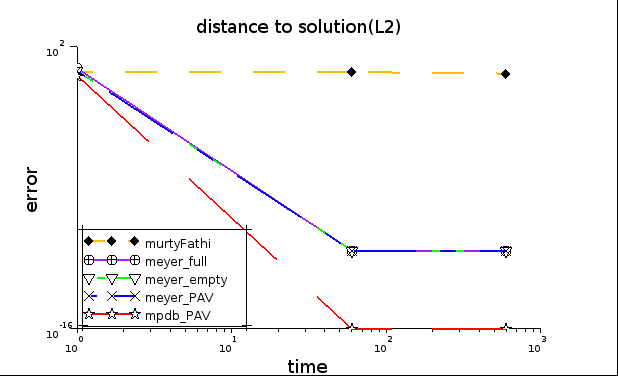}&
\includegraphics[width=70mm]{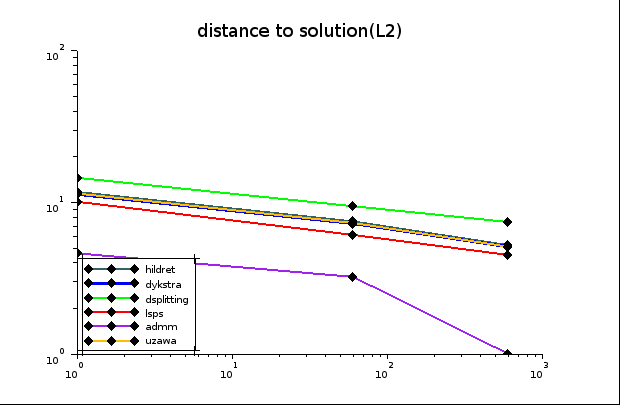}
\end{tabular}
\caption{Distance to the solution ($L_2$ norm) for a Gaussian white noise signal of size $500$. (Left) Algorithms with time-finite convergence: the best performance are achieved by MPDB with a PAV's inspired initialization. (Right) Algorithms with asymptotic convergence: no algorithm  converges.}
\label{size500L2S3}
\end{figure*}

\newpage
\onecolumn
\subsubsection{Comparative evaluations on signals of  size 1000}
\begin{figure*}[!ht]
\begin{tabular}{lllcc}
\includegraphics[width=70mm]{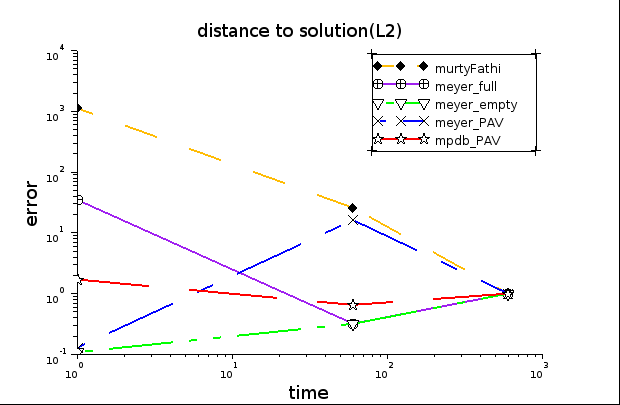}&
\includegraphics[width=70mm]{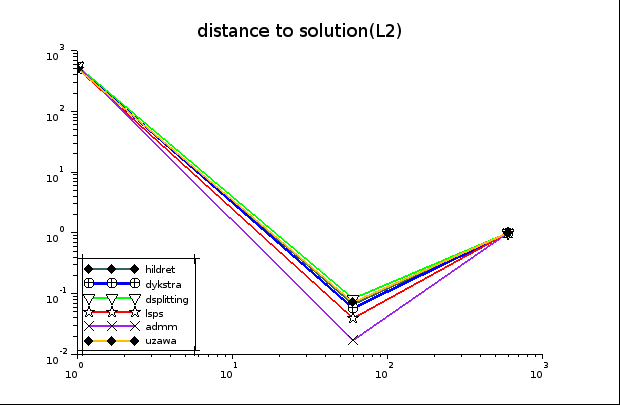}\\
\includegraphics[width=70mm]{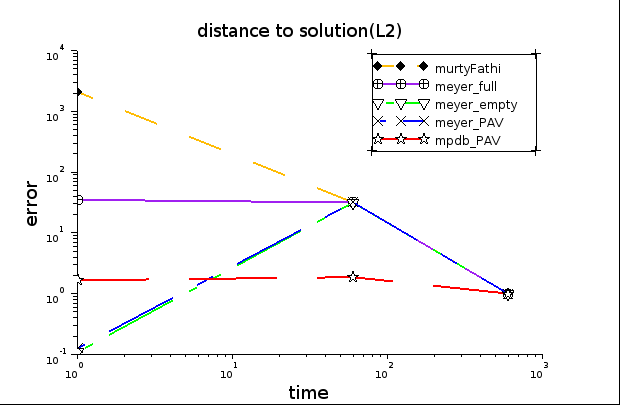}&
\includegraphics[width=70mm]{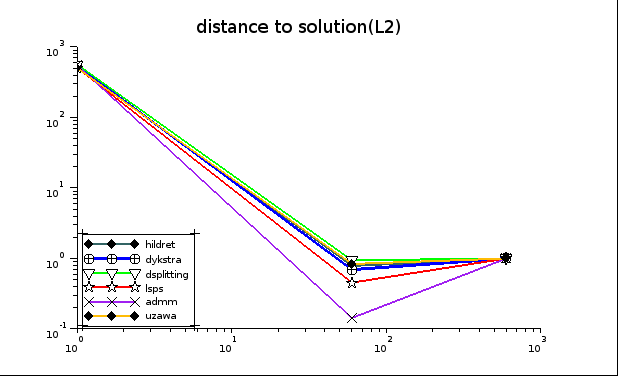}\\
\includegraphics[width=70mm]{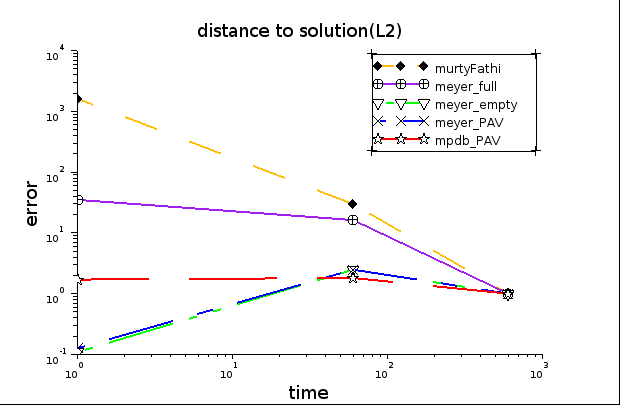}&
\includegraphics[width=70mm]{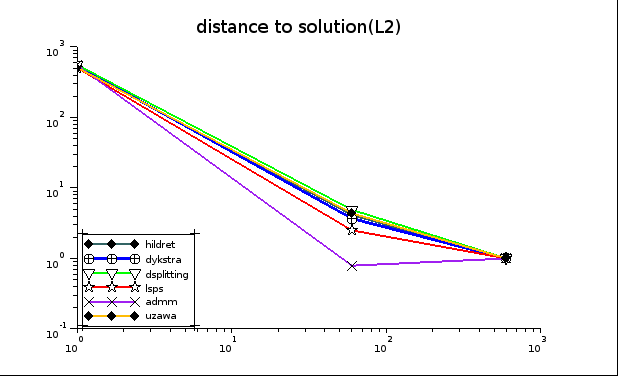}
\end{tabular}
\caption{Distance to the solution ($L_2$ norm) for a signal of type  $S_1$ of size $1000$. From up to the bottom, three increasing level of noise have been added. (Left) Algorithms with time-finite convergence: the best performance are achieved by MPDB with a PAV's inspired initialization. (Right) Algorithms with asymptotic convergence: no algorithm converges.}
\label{size1000L2S1}
\end{figure*}

\newpage 

\begin{figure*}
\begin{tabular}{llllcc}
\includegraphics[width=70mm]{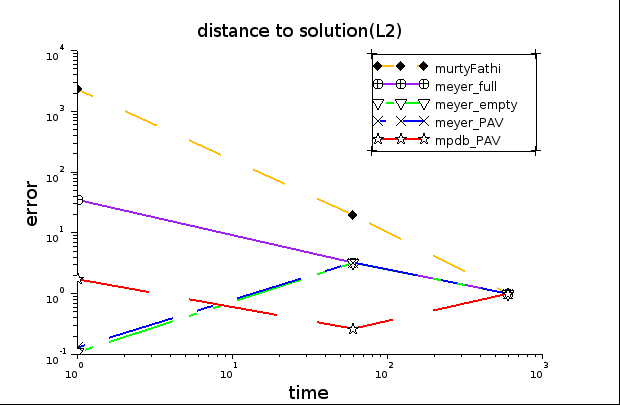}&
\includegraphics[width=70mm]{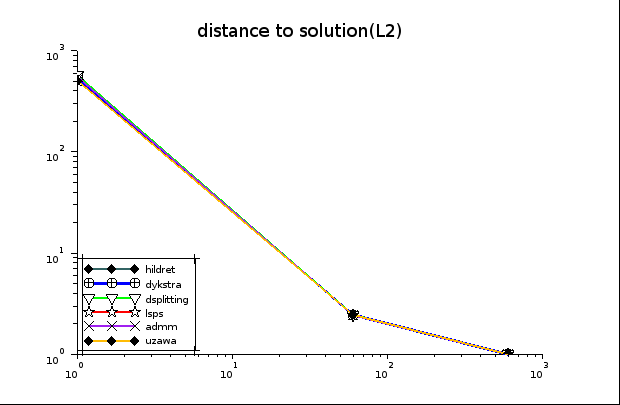}\\
\includegraphics[width=70mm]{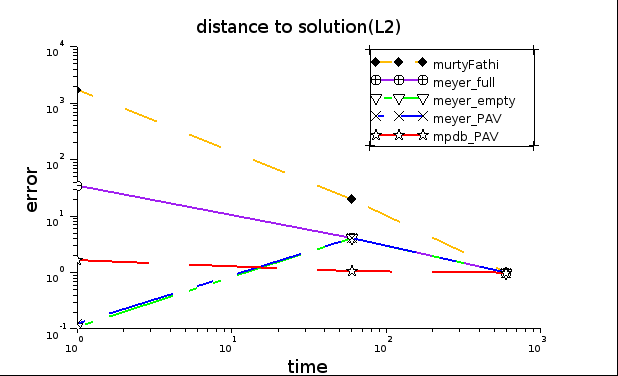}&
\includegraphics[width=70mm]{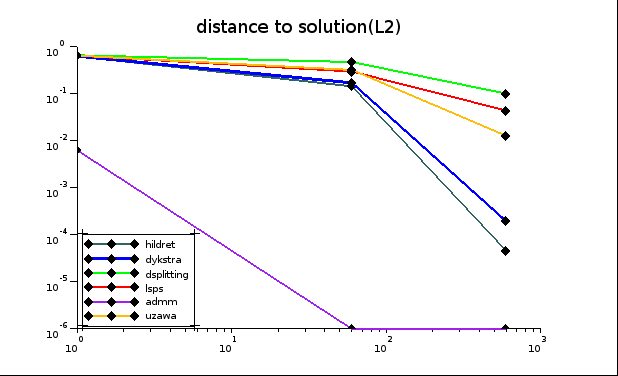}\\
\includegraphics[width=70mm]{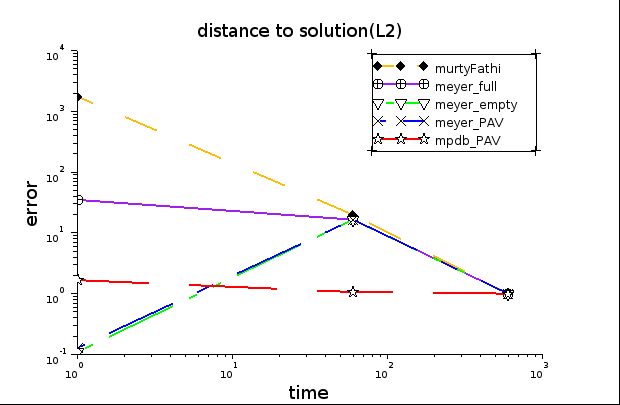}&
\includegraphics[width=70mm]{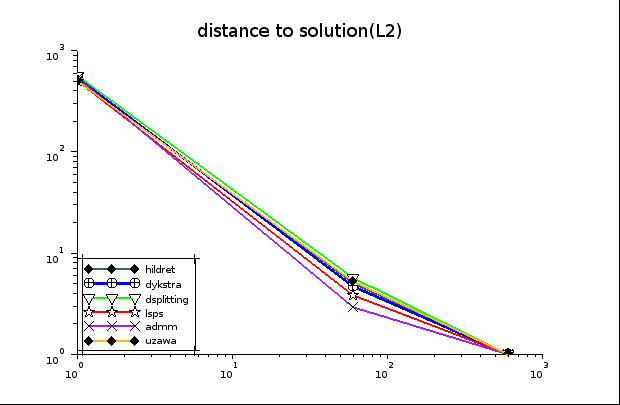}
\end{tabular}
\caption{Distance to the solution ($L_2$ norm) for a signal of type $S_2$ of size $1000$. From up to the bottom, three increasing level of noise have been added. (Left) Algorithms with time-finite convergence: the best performance are achieved by MPDB with a PAV's inspired initialization. (Right) Algorithms with asymptotic convergence: no algorithm converges.}
\label{size1000L2S2}
\end{figure*}

\newpage 

\begin{figure*}
\begin{tabular}{llllcc}
\includegraphics[width=70mm]{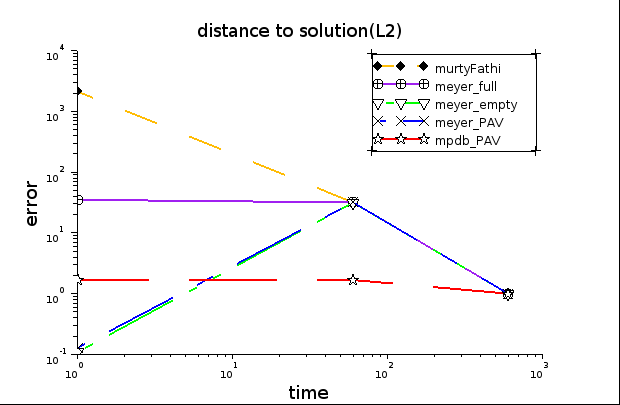}&
\includegraphics[width=70mm]{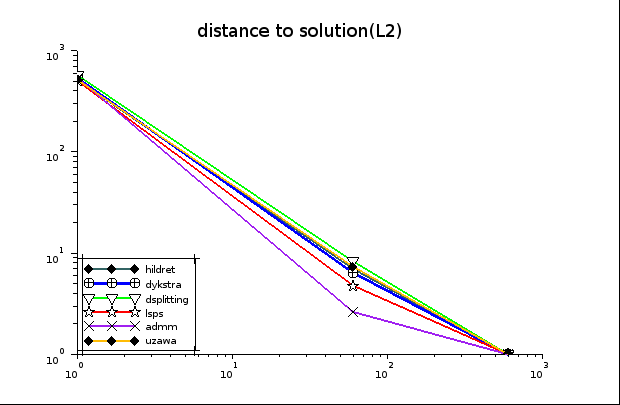}\\
\includegraphics[width=70mm]{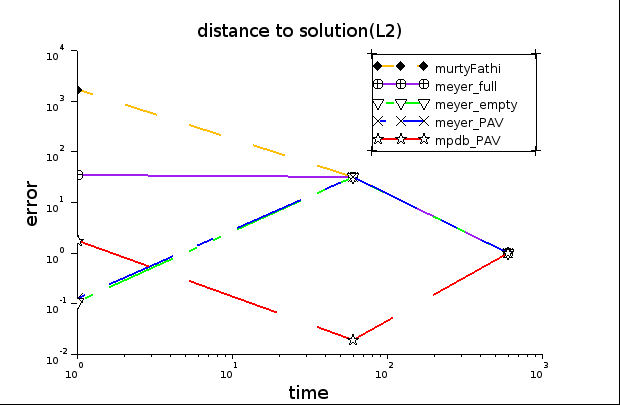}&
\includegraphics[width=70mm]{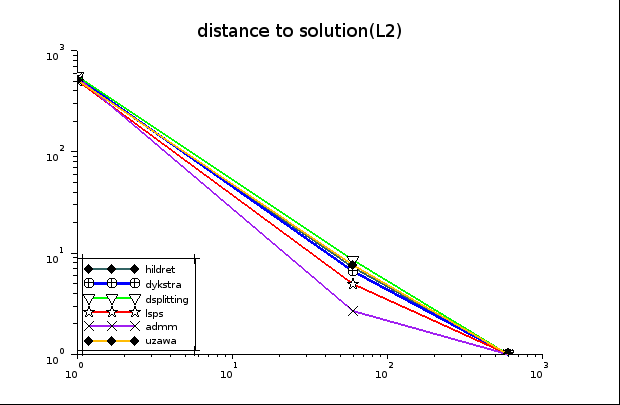}\\
\includegraphics[width=70mm]{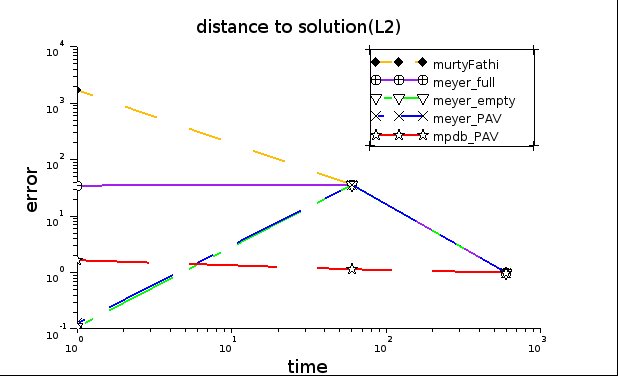}&
\includegraphics[width=70mm]{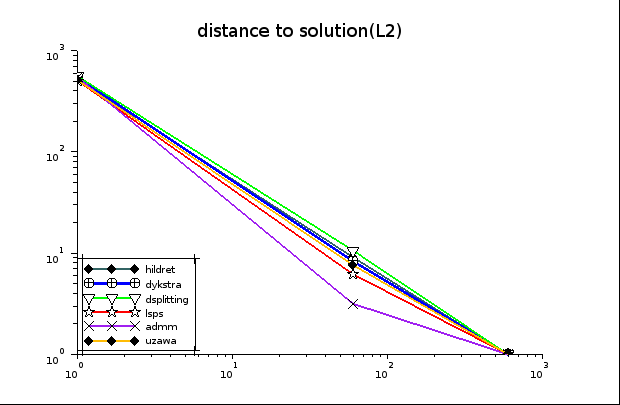}
\end{tabular}
\caption{Distance to the solution ($L_2$ norm) for a signal of type $S_3$  of size $1000$. (Left) Algorithms with time-finite convergence: the best performance are achieved by MPDB with a PAV's inspired initialization when the round-off error does not dominate the calculation and by the Meyer algorithm whose initial active set is empy. This is easy to understand since, for a pure noise signal an high degree of freedom is expected at the solution and therefore the empty active set is close to the the final active set. (Right) Algorithms with asymptotic  convergence: no algorithm converges.}
\label{size1000L2S3}
\end{figure*}

\clearpage 
\newpage
\lhead[\footnotesize\thepage\fancyplain{}\leftmark]{}\rhead[]{\fancyplain{}\rightmark\footnotesize\thepage}

\vskip 14pt
\noindent {\large\bf Supplementary Materials}

The online supplementary materials contain the pseudocode of the reviewed algorithms as well as their implementation in Scilab.
\par
\vskip 14pt
\noindent {\large\bf Acknowledgements}
The author would like to thank Lionel Moisan for insighful suggestions. This work has been carried out  during a postdoctoral stage in the Laboratory of Applied Mathematics (MAP5, CNRS UMR 8145) at Paris Descartes University. 
\par


\bibliography{biblio}

\end{document}